\documentclass[10pt]{article}
\newcommand\beq{\begin{equation}}
\newcommand\eeq{\end{equation}}
\newcommand\bea{\begin{eqnarray}}
\newcommand\eea{\end{eqnarray}}

 \let\mathfrak\undefined
\usepackage{amsmath,amssymb,graphicx,latexsym,eufrak,mathrsfs,bbm,parskip}
\begin{large}
\author{Yendrembam Chaoba Devi, Kumar Jang Bahadur Ghosh, Biswajit Chakraborty \\
S.N.Bose National Centre For Basic Sciences, JD Block, Salt Lake, Kolkata-700098, India\\
 and \\
Frederik G. Scholtz\\
Institute of Theoretical Physics, University of Stellenbosch, Stellenbosch 7600, South Africa, \\
National Institute for Theoretical Physics (NITheP), Stellenbosch 7600, South Africa}
\end{large}
\begin{LARGE}
\title{Thermal effective potential in two- and three- dimensional Non-commutative spaces}
\end{LARGE}

\begin{document}
\maketitle

\begin{abstract}
The issue of thermal correlation functions and the associated effective statistical potential in two-dimensional Moyal space, arising in the twisted approach to implement rotational symmetry, has been revisited in an operatorial formulation where no explicit star product is used initially. The corresponding results using Moyal and Voros star products are then easily obtained by taking the corresponding overlap with Moyal and Voros bases. in contrast to the Moyal case where the concept of distance and, in particular, the relative separation between a pair of particles remain ambiguous when the Moyal star product is used, the Voros basis is more physical and the inter-particle distance can be introduced unambiguously. The forms of the correlation function and the effective potential are found to be same as the Moyal case except that the thermal wavelength undergoes a non-commutative deformation, ensuring that it has a lower bound of the order of $\sqrt{\theta}$. It is shown that in a suitable basis (called here quasi-commutative basis) in the multi-particle sector the thermal correlation function coincides with the commutative result both in the Moyal and Voros cases along with the restoration of the Pauli principle, except that in the Voros case the thermal wavelength, again, gets a non-commutative correction. Finally, we extend our result to three-dimensional non-commutative space and compute the correlation function and effective potential using both twisted and quasi-commutative bases in the Moyal and Voros cases.  We find that there is $SO(3)\rightarrow SO(2)$ symmetry breaking in the effective potential, which also violates the Pauli principle, even for a pair of free particles, despite the fact that a deformed co-product is used to construct twisted symmetric/anti-symmetric basis. However, this $SO(3)$ symmetry, along with Pauli principle, is restored once we use  the quasi-commutative bases. 
\end{abstract}

\pagebreak
\section{Introduction}
It was realized sometimes back by Doplicher et.al \cite{b1} from the consideration of both general relativity and quantum mechanics that the localization of an event in space-time with arbitrary accuracy is operationally impossible and this feature is captured by postulating a non vanishing commutation relation between the coordinates which are now promoted to the level of operators. In its simplest form they are given as
\begin{equation}
[\hat{x}_\mu,\hat{x}_\nu] = i\theta_{\mu\nu}  \label{noncom.rel}
\end{equation}
where $\theta_{\mu\nu}$ is taken to be an antisymmetric matrix and its entries are viewed as new fundamental constants \cite{b9}. This form of non commutativity also follows from low energy limit of string theory \cite{b2}.\\

Analysis of quantum field theory in the background of such non-commutative spaces is expected to provide insight into the structure of quantum gravity as $\sqrt{\theta}$ is expected to be of the order of Planck length scale. Introducing such a length scale can have some serious consequences. For example, the structure of the commutation relation (\ref{noncom.rel}), with ~$\theta_{\mu\nu}$~ held fixed (i.e. not a tensor) signals the violation of Lorentz symmetry or simply the rotational $SO(3)$ symmetry in a non-relativistic system if the time \textquoteleft t' is taken to be the usual c-number parameter ($\theta_{0i}=0$), rather than an operator. As has been shown in \cite{b15}, this symmetry can be restored formally by deforming the co-product using the Drinfeld twist. This in turn implies, according to the approach followed in \cite{b14}, that the projection operator used to project multi-particle states into symmetric/antisymmetric subspaces to construct bosonic/fermionic states too should be deformed, 
thereby obtaining twisted bosons/fermions. In \cite{b11}, these twisted fermions were shown to violate Pauli exclusion principle by computing thermal correlation function for a pair of twisted fermions, although it preserves the Fermi Dirac statistics \cite{P.Basu}.
This computation, however, was carried out using Moyal star product, which is essentially associated with certain basis (called Moyal basis in \cite{b5}). In fact, in the operatorial formulation of non-commutative quantum mechanics involving Hilbert-Schmidt operator, which was initiated in \cite{b3, b4}, one can bypass the use of any star product and hence can avoid the use of any associated bases and any ambiguities that may result there from \cite{b5}. For example, in two-dimensional non-commutative Moyal plane one can introduce two types of canonical star products: Moyal and Voros and one can identify the respective Moyal and Voros basis so that the representation, in any of these bases, of a composite state(obtained by a simple operator multiplication of a pair of states) is the same as the one obtained 
by composing the respective representations of individual states by Moyal/Voros star products \cite{b5}. However, it is only the Voros basis which can be regarded as physical, as this conforms to POVM (Positive Operator Valued Measure), unlike the Moyal basis. In particular the Voros basis turns out to be a coherent state $|z)$ representing a maximally localized state in the non-commutative plane. Indeed, it has been shown recently \cite{b16} that one can compute the spectral distance, \textit{a la} Connes \cite{b17} between a pair of neighboring states 
$|z)_V$ and $|z+dz)_V$ to get a Euclidean geometry: $d^2(|z),|z+dz))=\frac{2\theta}{3}d\bar{z}dz.$ But such a distance function cannot be assigned between the pair of neighboring states $|\vec{x})_M$ and $|\vec{x}+d\vec{x})_M$ corresponding to the Moyal basis. The primary reason is that the Moyal basis does not conform to the requirement of POVM, as we have mentioned above. Besides, this Moyal basis turns out to be the eigenstate of commuting \textquotedblleft position-like \textquotedblright~ observables: 
$\hat{\vec{X}}^c|\vec{x})_M = \vec{x}|\vec{x})$, which are defined as the average of left and right actions of the noncommutative position operators \cite{Bal} or equivalently obtained by a suitable linear transformation in the phase space as 
\begin{equation}
 \hat{X}_i^c= \frac{1}{2}( \hat{X}_i^L + \hat{X}_i^R ) = \hat{X}_i+\frac{\theta}{2}\epsilon_{ij}\hat{P}_j, \label{transfm.}
\end{equation}
satisfying $[\hat{X}_1^c,\hat{X}_2^c]=0$ and therefore cannot be interpreted as the position observables \cite{b5}. Consequently, this Moyal basis is a purely mathematical construct and is devoid of any physical meaning.

These considerations therefore motivate us to have a re-look at this whole twisted formalism \cite{b14} completely at the operatorial level, observe more closely the principle(s) adopted here to enable us, to see whether this formalism where one constructs twisted bosons/fermions to accommodate deformed co-product, is an inevitable consequence of non-commutative space-time at least of the type given in (\ref{noncom.rel}). This is expected to shed light on other alternative formalisms  existing in the literature ( see for example \cite{b10} ). 
Indeed, the formalism adopted in \cite{b9} was used to compute the thermal correlation function in \cite{b13}. Carrying out this analysis at the level of non-commutative quantum mechanics automatically paves the way for introducing second quantized non-relativistic field operators, which has a built-in tensor-product structure, where the left slot corresponds to particle creation/annihilation operator and the right slot is a momentum eigenstate. Like the \textquotedblleft first quantized \textquotedblright~ quantum mechanical state, one can also obtain Moyal/Voros space representations of this abstract second-quantized field operators by taking appropriate overlap with M/V basis. With this, the stage is set to carry out the computations of the thermal correlation function in the Voros basis, where one can sensibly talk about the inter-particle distance. But, as has been discussed in \cite{b5}, the Voros wave functions $(z|\psi)$ of a state $|\psi)$, besides belonging to Schwartz class like its \
\textquotedblleft Moyalian\textquotedblright~ counterpart, need to satisfy an additional smoothness criterion at small length scale $\sim \sqrt{\theta}$; oscillators with wavelengths $\lesssim \sqrt{\theta}$ are suppressed exponentially. Since the thermal wavelength $\lambda =\sqrt{\frac{2\pi\beta}{m}}$ occurring in the expression of correlation function in \cite{b11}, computed in the Moyal basis which can be made smaller than $\lesssim \sqrt{\theta}$ for high enough temperature, one expects to find appropriate deformation in the corresponding result in Voros basis.

Secondly, as we show below, the twisted basis in the momentum space in the multi-particle sector is equivalent to a basis, up to an overall phase, which is symmetric/anti-symmetric under the usual exchange operation (i.e. not the deformed one) enabling one to define the usual bosons/fermions (and not their twisted counterparts), which nevertheless retains some deformations, stemming from non-commutativity. It will thus be interesting to study the structure of the thermal correlation function in these bases as well, which we henceforth refer to as \textquotedblleft quasi-commutative basis\textquotedblright,~ for reasons that will become clear in the sequel.

Finally, we would like to extend our computation from the planar case to the more realistic and physical Voros basis in odd three-dimensional space as well. Indeed in three-dimensional space such a Voros basis was introduced in \cite{b6} satisfying the over completeness relation as in the two-dimensional case. It is therefore imperative to check whether this too satisfies the POVM criterion and saturates the uncertainty relation both in 3D coordinate space and 6D phase space. Related to this, is the structure of 3D non-commutative parameter $\theta_{ij}$. Being singular, this matrix admits a vector $\vec{\theta}=\{\theta_i=\frac{1}{2}\epsilon_{ijk}\theta_{jk}\},$ dual to $\theta_{ij}$ and pointing in a particular direction which behaves like a commutative axis in a rotated frame. It thus spoils the isotropicity i.e. $SO(3)$ symmetry of 3D space. Despite restoring the $SO(3)$ symmetry through the deformed co-product at the level of commutator as in \cite{b15}, it was shown in \cite{b6} to violate $SO(3)$ 
symmetry at the level of the action in presence of interaction. It will therefore be quite interesting to look for any signature of the violation of this symmetry for a system containing a pair of free particles, itself. As a thermal effect the structure of the 3D correlation function or the resulting statistical potential should tell us immediately about this violation of $SO(3)$ symmetry. Yet another place where this violation can also show up, is the structure of the variance matrix which occurs in the symplectic invariant formulation of the uncertainty relation, which we use here. Since, however, the Williamson's theorem \cite{b7} and the associated technique for symplectic diagonalization \cite{b8} is not known to hold in our context, we can make use of the transformation (\ref{transfm.}) to obtain the corresponding commutative variance matrix and its symplectic spectrum and try to see whether the saturation condition holds for the entire variance matrix and also for each distinct \textquotedblleft 
modes \textquotedblright~ 
(which are now de-coupled from each other) in the same manner. Finally,the resulting profile of the thermal effective potential may be used to study the nature of the violation of Pauli exclusion principle, if any. This is expected to pave the way to study the astrophysical implications.

The plan of the paper is given in the following manner: In section II, we review the formulation of non-commutative quantum mechanics on two-dimension and three-dimension. In section III, we provide a symplectic invariant formulation of the uncertainty relation. In section IV, we compute the non-commutative Variance matrix for Voros basis in three-dimension and find that Voros basis states are maximally localized in phase space although it does not represent a maximally localized state in 3D non-commutative space. This single particle formulation of non-commutative quantum mechanics is extended in section V to two-particle system, where we try to formulate the twisted symmetric/anti-symmetric of \cite{b14} through our operatorial approach. 
In section VI, the twisted formulation has been extended further for more than two particles and here we have introduced a \textquotedblleft quasi-commutative\textquotedblright~  momentum basis which is the usual symmetric/antisymmetric basis differing from the twisted basis by only a phase factor in the momentum space. With this, in section VII, we have discussed the second quantization through the second quantized creation/annihilation operator so as to establish a contact with non-relativistic quantum field theory. Introducing the creation and annihilation operators in these two types of multi-particle momentum bases in section VII, we have defined the abstract field operators in section VIII and discuss its action on arbitrary state in position and momentum representations. Then in section IX, we compute the two-particle correlation functions for a free gas in both two and three-dimensions and represent them in both Moyal and Voros bases using both twisted and quasi-commutative bases and obtain the 
corresponding effective potentials which is 
eventually plotted with respect to the inter-particle distance. Finally, in section X, we conclude the paper.

\section{Formulation of Non-commutative quantum mechanics in 2-D and 3-D spaces}
\label{section2}
 In two dimension, the non-commutative Heisenberg algebra(in the unit $\hbar =1$) can be written as
\begin{equation}
 \left[ \hat{x}_i , \hat{x}_j \right] = i \theta _{ij}=i\theta\epsilon_{ij},~~~ \left[ \hat{x}_i , \hat{p}_j \right] = i \delta _{ij} ~~~\text{and}~~~\left[ \hat{p}_i , \hat{p}_j \right] = 0 ~~~\forall i,j= 1,2
\end{equation}
 Defining the creation ~$b^\dagger =\frac{1}{2\theta}\left( \hat{x}_1 - i \hat{x}_2\right)$~ and annihilation operators $~b= \frac{1}{2\theta}\left( \hat{x}_1 + i \hat{x}_2\right)~$ satisfying ~$[b, b^\dagger]=1$~, the non-commutative two dimensional classical configuration space can be written as
 \begin{equation}
\mathcal{H}_c = \text{span} \lbrace \mid n\rangle \rbrace _{n=0}^{\infty} \label{configuration}
\end{equation}
where ~$| n \rangle= \frac{1}{\sqrt{n!}}(b^\dagger)^n|0\rangle$~ is the eigenstate of the operator ~$b^\dagger b$~:~($b^\dagger b|n\rangle=n|n\rangle $)\\

The corresponding quantum Hilbert space, the elements of which represent the physical states, can then be taken as the set of Hilbert- Schmidt operators which are all bounded trace-class operators over $\mathcal{H}_c$
\begin{equation}
\mathcal{H} _q = \lbrace \psi : \text{tr}_c ( \psi^{\dagger} \psi)< \infty \rbrace 
\end{equation}
The elements of $\mathcal{H}_q$ are denoted by a round bracket ~$|\psi)$~ and the inner product between them is defined as 
\begin{equation}
 (\phi|\psi) = \text{tr}_c(\phi^\dag \psi)
\end{equation}
where the subscript $c$ refers to tracing over $\mathcal{H}_c$ and $\dag$ denotes Hermitian conjugation on $\mathcal{H}_c$ while $\ddagger$ will denotes the same on $\mathcal{H}_q$.

If $\hat{X}_i$ and $\hat{P}_i$ are the representations of the operators $\hat{x}_i$ and $ \hat{p}_i$ respectively acting on $\mathcal{H}_q$, then a unitary representation is obtained by the following action:
\begin{equation}
 \hat{X}_i \psi = \hat{x}_i \psi, ~~~ \hat{P}_i \psi = \frac{1}{\theta}\epsilon_{ij}[\hat{x}_j, \psi]
\end{equation}
It is easily verified that the momentum eigenstates $|\vec{p})$ are given by
\begin{equation}
 |\vec{p}) = \sqrt{\frac{\theta}{2\pi}} e^{i\vec{p}.\hat{\vec{x}}}, ~~~i.e.~ \hat{P}_i |\vec{p}) = p_i |\vec{p})
\end{equation}
and they satisfy the usual orthonormality condition and resolution of identity
\begin{equation}
(\vec{p}|\vec{p}') = \delta^2(\vec{p}- \vec{p}'),~~~~~  \int d^2p |\vec{p})(\vec{p}| = 1_q.
\end{equation} 
One can then introduce the Voros basis \cite{b5}
\begin{equation}
 |\vec{x} )_V = \sqrt{\frac{\theta}{2\pi}} \int d^2p~e^{-\frac{\theta p^2}{4}}~e^{-i\vec{p}.\vec{x}}~|\vec{p}) \label{vorosbasis}
\end{equation}
and the Moyal basis as
\begin{equation}
|\vec{x})_M = \int \frac{d^2p}{2\pi} ~e^{-i\vec{p}.\vec{x}}~|\vec{p}). \label{Moyal}
\end{equation}

As the Voros basis can also be written as $ ~|\vec{x})_V = |z,\bar{z})=|z\rangle\langle z|$, in terms of the coherent states $|z\rangle = e^{|z|^2/2}e^{zb^{\dagger}}|0\rangle$, this represents the maximally localized state. Further,the representation of any composite state
~$ |\psi\phi)$~ in this Voros/Moyal basis automatically yields the Voros/Moyal star product composed expression of the corresponding representations of the individual states:
\begin{equation}
 _{V/M}(\vec{x}|\psi\phi) = \lambda_{V/M}~_{V/M}(\vec{x}|\psi)*_{V/M} {_{V/M}(\vec{x}|\phi)}  \label{composed}
\end{equation}
with ~$\lambda_V= 4\pi^2$~ and ~$\lambda_M=\sqrt{2\pi\theta}$, the Voros basis was shown to conform to the requirement of POVM, in contrast to the Moyal basis \cite{b5} as one can see that the resolution of identity for the Voros basis 
\begin{equation}
1_q = \int d^2z |\vec{x})_V *_V {_V(\vec{x}|}
\end{equation}
has the integrand $\pi_z=|\vec{x})_V *_V {_V(\vec{x}|}$, which is a positive, but non-orthogonal unnormalized projection operator
\begin{equation}
\int dx\; dy\;\pi_z=1_q\,\quad (\psi|\pi_z|\psi)\geq 0,\forall\psi\,,\quad \pi_z\pi_w\ne \delta(z-w)\,,\quad \pi_z^2\propto \pi_z.
\end{equation} 
 
In $3-D$, the algebra satisfied by the coordinate operators can be written as
\begin{equation}
\left[ \hat{x}_i , \hat{x}_j \right] = i \theta _{ij} = i\epsilon _{ijk} \theta _k , ~~ i,j,k = 1,2,3 
\end{equation}
where $\theta _{ij}$ is a $3\times 3$ anti symmetric matrix and $ \vec{\theta} = \{\theta_k\} $ is a vector dual to this.\\

By formally transforming the coordinate system $ (\hat{x}_i \rightarrow \hat{\bar{x}}_i)$ by a suitable SO(3) rotation say $\bar{R}$, we can orient the vector $\vec{\theta}$ in the fiducial frame along say the third axis. For example, if
\begin{equation}
 \vec{\theta} = \theta\begin{pmatrix}
                      \sin\alpha\cos\beta\\
                      \sin\alpha\sin\beta\\
                      \cos\alpha\\
                      \end{pmatrix} \label{theta}
\end{equation}
then the transformation   ~~~$\hat{x}_i\rightarrow \hat{\bar{x}}_i = \bar{R}_{ij}\hat{x}_j$~~~performed with
\begin{equation}
 \bar{R}=\begin{pmatrix}
          \cos\alpha\cos\beta & \cos\alpha\sin\beta & -\sin\alpha \\
          -\sin\beta & \cos\beta & 0 \\
          \sin\alpha\cos\beta & \sin\alpha\sin\beta & \cos\alpha \\ 
        \end{pmatrix}     ~~~~~\in SO(3)   \label{rotmatrix} 
\end{equation}

reduces the  non-commutative coordinate algebra in barred three dimensional frame to
\begin{eqnarray}
\left[ \hat{\bar{x}}_1 , \hat{\bar{x}}_2 \right] &=& i \theta \\  
 \left[ \hat{\bar{x}}_\alpha , \hat{\bar{x}}_3 \right] &=& 0 ,~~~~~~~ \alpha= 1,2
\end{eqnarray}
 This implies that the non-commutative $3D$ classical configuration space can be constructed as a tensor product space of the $2D$ classical configuration space and one-dimensional Hilbert space spanned by the eigenstates of $\hat{\bar{x}}_3$.
\begin{equation}
i,e ~~~ \mathcal{H} _c = \text{span} \lbrace \mid n , \bar{x}_3\rangle   \rbrace 
\end{equation}
The corresponding Quantum Hilbert space can be identified as 
\begin{equation}
\mathcal{H} _q = \left\lbrace  \psi(\hat{\bar{x}}_i) : \left[ \hat{\bar{x}}_3 , \psi \right] =0;~\int \frac{d\bar{x}_3}{\sqrt{\theta}}\text{tr}_c ^{\prime} ( \psi^{\dagger} \psi)< \infty \right\rbrace
\end{equation}
where tr$_c ^{\prime}$ denotes the restricted trace over the non-commutative $2D$ plane.\\

 This means that the elements of $\mathcal{H} _q$ are the Hilbert-Schmidt operators on ~$\mathcal{H} _c$~ and which satisfy the additional constraint ~$[\hat{\bar{x}}_3 , \psi]=0$~ and the inner product between these elements is defined as usual
\begin{equation}
 (\phi|\psi) = \text{tr}_c (\phi^\dag \psi) = \int \frac{d\bar{x}_3}{\sqrt{\theta}}~ \text{tr}'_c(\phi^ \dag \psi) 
\end{equation}
To define the action of the momentum operators on the quantum Hilbert space, it is convenient to introduce a further coordinate \textquoteleft $\hat{\bar{x}}_4$' such that
\begin{equation}
 [\hat{\bar{x}}_j, \hat{\bar{x}}_4] = i\theta \delta_{j3}
\end{equation}
i.e. $\hat{\bar{x}}_4$ commutes with $\hat{\bar{x}}_1, \hat{\bar{x}}_2$ and is conjugate to $\hat{\bar{x}}_3$ so that $\hat{\bar{x}}_4 = -i\theta \frac{\partial}{\partial \hat{\bar{x}}_3}$. Then the action of the momentum operators in the barred frame on the quantum Hilbert space can be expressed through the adjoint action:
\begin{equation}
 \hat{\bar{P}}_\mu \psi = \frac{1}{\theta} \Gamma _{\mu\nu} [\hat{\bar{x}}_\nu, \psi]; ~~~\mu,\nu = 1,2,3,4,
\end{equation}
where
\begin{equation}
 \Gamma = \begin{pmatrix}
           0 & 1 & 0 & 0 \\
           -1 & 0 & 0 & 0 \\
           0 & 0 & 0 & 1 \\
           0 & 0 & -1 & 0 \\
          \end{pmatrix}
\end{equation}
By the constraint $[\hat{\bar{x}}_3,\psi]=0$, we have 
\begin{equation}
 \hat{\bar{P}}_4 \psi = 0
\end{equation}
so there are only three non-trivial momenta.

Then the action of the components of momenta in the original frame can be obtained through linearity as
\begin{eqnarray}
\nonumber \hat{P}_i \psi &=& (\bar{R}^{-1})_{ij} \hat{\bar{P}}_j \psi \\ 
&=& \frac{1}{\theta} (\bar{R}^{-1})_{ij} \Gamma_{j\mu} [\hat{\bar{x}}_\mu, \psi]
 \end{eqnarray}
We can verify that the simultaneous eigenstates of the above commuting momentum operators are given by
\begin{equation}
 |\vec{p}) = \frac{\theta^{\frac{3}{4}}}{2\pi} ~e^{ip_i\hat{x}_i} = \frac{\theta^{\frac{3}{4}}}{2\pi} ~e^{i\bar{p}_3\hat{\bar{x}}_3}~e^{i\bar{p}_\alpha\hat{\bar{x}}_\alpha};~~~ \hat{P}_i |\vec{p} ) = p_i |\vec{p}).
\end{equation} 
Note that $\vec{p}.\hat{\vec{x}}$ is a scalar under an $SO(3)$ rotation. These momentum eigenstates will play an important role in the quantum Hilbert space. In complete analogy with the 2-D case, these states too satisfy the orthonormality and the completeness relations
\begin{equation}
(\vec{p}'|\vec{p}) = \delta^3(\vec{p}' - \vec{p});~~~~~\int d^3p ~|\vec{p})(\vec{p}| = 1_q.
\end{equation}
Here also, we can define the Voros basis as 
\begin{equation}
 |\vec{x})_V = \frac{\theta^{\frac{3}{4}}}{\sqrt{2\pi}} \int d^3p ~e^{-\frac{\theta p^2}{4}}e^{-i\vec{p}.\vec{x}}~|\vec{p}) \label{voros}
\end{equation} which satisfies the completeness relation, if composed through Voros star product:
\begin{equation}
 \int \frac{d^3x}{(2\pi)^2\theta^\frac{3}{2}}|\vec{x})_V*_V{_V(\vec{x}|}=1_q
\end{equation}
where \begin{equation}
 *_V =  e^{\frac{i}{2}\theta_{ij}^V \overleftarrow{\partial}_i \overrightarrow{\partial}_j} 
\end{equation} with ~$\theta_{ij}^V =-i\theta\delta_{ij}+\theta_{ij} $,~ and have non-orthogonal overlap between any pair of such states:
\begin{equation}
 _V(\vec{x}'|\vec{x})_V = \sqrt{2\pi}~e^{-\frac{1}{2\theta}(\vec{x}'-\vec{x})^2}
\end{equation}
The corresponding overlap of this basis with momentum basis is 
\begin{equation}
 _V(\vec{x}|\vec{p}) = \frac{\theta^\frac{3}{4}}{\sqrt{2\pi}}~e^{-\frac{\theta p^2}{4}}~e^{i\vec{p}.\vec{x}}
\end{equation}
Next, we define the 3D Moyal basis (the counterpart of (\ref{Moyal}))as
\begin{equation}
 |\vec{x})_M = \int \frac{d^3p}{(2\pi)^\frac{3}{2}} e^{-i\vec{p}.\vec{x}}|\vec{p})
\end{equation}
which also satisfies the completeness relation
\begin{equation}
 \int d^3x |\vec{x})_M {*_M} _M(\vec{x}| =  \int d^3x |\vec{x})_M  {_M(\vec{x}|} = 1_q,
\end{equation}
where \begin{equation}
       *_M = e^{\frac{i}{2}\theta^M_{ij}\overleftarrow{\partial_i}\overrightarrow{\partial_j}}, ~~~\text{with}~~~\theta^M_{ij} = \theta_{ij}=\theta \epsilon_{ij}.
      \end{equation}
The Moyal basis is an orthonormal basis 
\begin{equation}
 _M(\vec{x}|\vec{x'})_M =\delta^3(\vec{x}-\vec{x'}),
\end{equation}
and its overlap with momentum basis is 
\begin{equation}
 (\vec{p}|\vec{x})_M =\frac{1}{(2\pi)^\frac{3}{2}} e^{-i\vec{p}.\vec{x}}.
\end{equation}
These Moyal basis states are the simultaneous eigenstates of $\hat{X}^c_i$ given by (\ref{transfm.}).

On the 3-D quantum Hilbert space $\mathcal{H}^{(3)}_q$ we can impose the additional structure of an algebra by defining the multiplication map \cite{b6}:
\begin{equation}
 m(|\psi)\otimes|\phi)) = |\psi\phi).
\end{equation}
such that expanding a generic state $|\psi)$  in terms of momentum eigenstates and going to barred frame:
\begin{equation}
 |\psi) = \frac{\theta^\frac{3}{4}}{2\pi}\int\frac{d^3p}{(2\pi)^\frac{3}{2}}~\psi(\vec{p})~e^{ip_i\hat{x}_i} = \frac{\theta^\frac{3}{4}}{2\pi}\int\frac{d^3\bar{p}}{(2\pi)^\frac{3}{2}}~\psi(\vec{\bar{p}})~e^{i\bar{p}_i\hat{\bar{x}}_i}
\end{equation}we can now prove the following composition rules (3D counterpart of eqn.(\ref{composed}))
\begin{equation}
 _{M/V}(\vec{x}|\psi\phi) = {\lambda_{M/V}} _{M/V}(\vec{x}|\psi)*_{M/V} {_{M/V}(\vec{x}|\phi)}, ~~~\text{with}~~~\lambda_V=1,~~~\lambda_M = 2\pi\theta^\frac{3}{4}
\end{equation}where \begin{equation}
                     _V(\vec{x}|\psi) = \frac{\theta^\frac{3}{4}}{\sqrt{2\pi}}\int \frac{d^3p}{(2\pi)^\frac{3}{2}}~\psi(\vec{p})~e^{-\frac{\theta p^2}{4}}~e^{ip_ix_i};~~~ _M(\vec{x}|\psi) = \int \frac{d^3p}{(2\pi)^\frac{3}{2}}~\psi(\vec{p})~e^{ip_ix_i} 
                    \end{equation}
But here we loose any obvious connection of the Voros basis to the coherent state. We can now investigate whether the uncertainty relation saturates for both phase space variables and also for just position coordinates for the Voros basis (\ref{voros}).\\

\section{A brief review of Robertson and Schr\"{o}dinger Uncertainty relations, Variance matrix and symplectic formalism in commutative ($\theta=0$) quantum mechanics}
The standard deviation $\Delta \hat{A}$ of any Hermitian operator $\hat{A}$ in a state $|\Psi\rangle$ can be written as
\begin{equation}
 \Delta \hat{A}^2  = \langle\Psi |(\hat{A}-\langle \hat{A}\rangle)^2|\Psi\rangle= \langle (\hat{A}-\langle \hat{A}\rangle)\Psi|(\hat{A}-\langle \hat{A}\rangle)\Psi\rangle =\langle f_A |f_A\rangle
\end{equation}
where \begin{equation}
     |f_A\rangle =|(\hat{A}-\langle \hat{A}\rangle)\Psi\rangle 
    \end{equation}

Then using the Schwarz inequality for a pair of such observables $\hat{A}$ and $\hat{B}$, we can write  
\begin{equation}
\Delta \hat{A}^2\Delta \hat{B}^2=\langle f_A |f_A\rangle\langle f_B|f_B\rangle\geq |\langle f_A|f_B\rangle|^2
\end{equation}

Splitting into real and imaginary terms, we can write
\begin{equation}
  |\langle f_A|f_B\rangle|^2 =\left(\frac{\langle f_A|f_B\rangle+\langle f_B|f_A\rangle}{2}\right)^2 + \left(\frac{\langle f_A|f_B\rangle-\langle f_B|f_A\rangle}{2i}\right)^2  \label{both}
 \end{equation}
Now using the fact that $\langle f_A|f_B\rangle =\langle \hat{A}\hat{B}\rangle -\langle \hat{A}\rangle\langle \hat{B}\rangle$ we see that ~$\langle f_A|f_B\rangle - \langle f_B|f_A\rangle = \langle [\hat{A}, \hat{B}]\rangle$~involves commutator whereas ~$\langle f_A|f_B\rangle + \langle f_B|f_A\rangle=\langle \{\hat{A},\hat{B}\}\rangle -2\langle \hat{A}\rangle\langle\hat{B}\rangle$~
involves anti-commutator.\\

We now introduce Robertson and Schr\"{o}dinger uncertainty one by one:\\

\underline{1. Robertson Uncertainty Relation}: Here we ignore the square of real part, to write \begin{equation}
                                                     |\langle f_A|f_B\rangle|^2 \geq (Im.\langle f_A|f_B\rangle)^2 = \left(\frac{\langle f_A|f_B\rangle-\langle f_B|f_A\rangle}{2i}\right)^2
                                                    \end{equation}
This gives the Robertson Uncertainty Relation
\begin{equation}
\Delta \hat{A}\Delta \hat{B} \geq \frac{1}{2i}\langle [\hat{A},\hat{B}]\rangle
\end{equation}
\underline{2. Schr\"{o}dinger Uncertainty Relation}: Here we retain both the squares of real and imaginary parts, to get, using (\ref{both})  
\begin{equation}
   |\langle f_A|f_B\rangle|^2 = \left(\frac{\langle f_A|f_B\rangle+\langle f_B|f_A\rangle}{2}\right)^2 + \left(\frac{\langle f_A|f_B\rangle-\langle f_B|f_A\rangle}{2i}\right)^2=\left(\frac{1}{2}\langle\{\hat{A},\hat{B}\}\rangle- \langle \hat{A}\rangle\langle \hat{B}\rangle\right)^2 + \left(\frac{1}{2i}\langle [\hat{A},\hat{B}]\rangle\right)^2                                                                 
\end{equation}
This finally gives the Schr\"{o}dinger Uncertainty Relation
\begin{equation}
 \Delta \hat{A}\Delta \hat{B} \geq\sqrt{\left(\frac{1}{2}\langle\{\hat{A},\hat{B}\}\rangle- \langle  \hat{A}\rangle\langle  \hat{B} \rangle\right)^2 + \left(\frac{1}{2i}\langle [\hat{A},\hat{B}]\rangle\right)^2}
\end{equation}
 Before proceeding further with the computation in the non-commutative case ($\theta\neq0$), let us discuss how this Schr\"{o}dinger's form of uncertainty relation can be recast in terms of Variance matrix. For this, let us rename the space and momenta operators by a single phase-space operator $\hat{Z}$ where $\hat{Z}_i =\hat{X}_i$ with $i=1,2,3.$ and  $\hat{Z}_{i+3}=\hat{P}_i$. Then the Schr\"{o}dinger uncertainty relation for the phase-space operators will be given by
\begin{equation}
 \Delta\hat{Z}_\mu\Delta\hat{Z}_\nu \geq\sqrt{\left(\frac{1}{2}\langle\{\hat{Z}_\mu,\hat{Z}_\nu\}\rangle- \langle \hat{Z}_\mu\rangle\langle \hat{Z}_\nu\rangle\right)^2 + \left(\frac{1}{2i}\langle [\hat{Z}_\mu,\hat{Z}_\nu]\rangle\right)^2} \label{sch}
\end{equation}
where the expectation values are to be taken in a certain state $|\Psi\rangle$ with $\mu,\nu =1,2,...6$.\\

The first squared term in the RHS of the above relation is the square of $\mu\nu-$ th element of the Variance matrix $V^0=\lbrace {V^0}_{\mu\nu}\rbrace$, where
\begin{equation}
 V^0_{\mu\nu}= \frac{1}{2}\langle\{\hat{Z}_\mu,\hat{Z}_\nu\}\rangle-\langle\hat{Z}_\mu\rangle\langle\hat{Z}_\nu\rangle
\end{equation}
and the second squared term can be identified as the square of $\mu\nu-$ th element of the symplectic matrix 
\begin{equation}
     \Omega^0= \lbrace\Omega^0 _{\mu\nu}\rbrace = \lbrace \frac{1}{2i}[\hat{Z}_\mu , \hat{Z}_\nu] \rbrace =\frac{1}{2}\begin{bmatrix}
                            0&0&0&1&0&0\\
                            0&0&0&0&1&0\\
                            0&0&0&0&0&1\\
                            -1&0&0&0&0&0\\
                            0&-1&0&0&0&0\\
                            0&0&-1&0&0&0\\
                           \end{bmatrix}
    \end{equation}
In general , the Variance matrix $V$ will be $2n\times2n$ square matrix corresponding to a $2n-$ dimensional phase space(in our case, $n=3$). This provides an exhaustive characterization of any Gaussian state. By Williamson's theorem \cite{b7} there exists a symplectic transformation $S \in Sp(2n, R)$ so that any arbitrary Variance matrix $V^0$ can be brought to a diagonal form $V^d = SV^0S^T$ ; $\Omega^0= S\Omega^0S^T$, where $V^d= \text{diag}(\nu_1/2,...,\nu_n/2,\nu_1/2, ...,\nu_n/2)$ comprises of the (up to the orderings of $\nu _j$) the symplectic eigenvalues of $V^0$, which are at least doubly degenerate. This symplectic spectrum is not to be confused with the ordinary spectrum, which is obtained by a similarity transformation. Rather, the symplectic spectrum can be obtained through the ordinary spectrum of $|2i\Omega^0V^0|$, as the composite object ($\Omega^0V^0$) undergoes a similarity transformation, if $V^0$ undergoes a symplectic transformation \cite{b8}. Correspondingly, the density matrix 
$\rho=|\Psi\rangle\langle\Psi|$ transforms a $\rho \rightarrow U(S) \rho U^{\dagger}(S)$,
where $\hat{U}(S)$ is a unitary operator implementing the symplectic transformation. We can notice, at this stage, that in this diagonal form, the $2n-$dimensional phase space splits into  n-copies of independent $2-$dimensional phase space. It is therefore convenient to consider the Schr\"{o}dinger's uncertainty relation (\ref{sch}) for 2D phase space re-written as, 
\begin{equation}
\Delta \hat{Z}_{\mu}\Delta \hat{Z}_{\nu} = \sqrt{(V^0_{\mu\nu})^2+(\Omega_{\mu\nu})^2} ~ ~ ~~~~\text{with}~ \mu, \nu = 1,2 ~ ~ (\text{no sum on} ~\mu , \nu)
\end{equation}
Identifying $\hat{Z}_1=\hat{X}_1$ and $\hat{Z}_4=\hat{P}_1$, this inequality is equivalent to $\Delta \hat{X}_1 ^2 \geq \frac{\nu_1}{2}$ ; $\Delta \hat{P}_1 ^2 \geq \frac{\nu_1}{2}$ whereas $\Delta \hat{X}_1 \Delta \hat{P}_1 \geq \frac{1}{2}$\\
Here we have taken $V_{11} ^d = V_{22} ^d = \frac{\nu_1}{2}$ and $V_{12} ^d = V_{21} ^d = 0$ without loss of generality, so that the spread $\Delta \hat{X}_1$ and $ \Delta \hat{P}_1$ are equal. Further, we have used the symplectic invariant form of $\Omega^0 :~ \Omega^0 _{12}=-\Omega^0 _{21}=\frac{1}{2}$ and $\Omega^0 _{11} = \Omega^0 _{22} = 0$. Compatibility among these three inequalities implies 
\begin{equation}
\nu_1 \geq 1
\end{equation}
or equivalently 
\begin{equation}
\text{det}~ V^0 \geq \frac{1}{4}. 
\end{equation}
Thus for a bonafide Variance marix $V^0$ we must have the symplectic spectrum to be such that $\nu_j \geq 1 ~ \forall j$ or more generally 
\begin{equation}
\text{det}~ V^0 \geq \frac{1}{4^n} \label{satcon}
\end{equation}
for the general ($2n$)- dimensional phase space. This provides a symplectic $Sp(2n,R)$ invariant formulation of the uncertainty relation. Finally, note that both Robertson and Schr\"{o}dinger form of uncertainty relations become equivalent in this diagonal form.

\section{Computation of Variance matrix in the non-commutative case for Voros basis in 3-D and saturation condition}
In our $3D$ non-commutative quantum system ($\theta \neq 0$), however, this formalism is not directly applicable. In particular, it is not known whether the Williamson's theorem, remains valid or not, in this context. We will therefore transform this non-commutative Variance matrix into a commutative one by using the transformation (\ref{transfm.}) and apply this formalism to show that for $3D$ Voros basis the symplectic invariant form of the uncertainty relation (\ref{satcon}) is indeed saturated indicating that like $2D$ case, the $3D$ Voros basis also corresponds to a maximally localized state in phase space. Starting with the normalized version of the Voros states (\ref{voros})
\begin{equation}
 |\vec{x})_V = \frac{\theta^\frac{3}{4}}{(2\pi)^\frac{3}{4}} \int d^3p ~e^{-\frac{\theta p^2}{4}}e^{-i\vec{p}.\vec{x}}~|\vec{p});~~~_V(\vec{x}|\vec{x})_V=1,
\end{equation}
we can now find the expectation values of $\hat{Z}_\mu$ and the composite $\hat{Z}_\mu\hat{Z}_\nu$ in the above normalized Voros states.
Now, the expectation value of $\hat{X}_i$ in these Voros states can be rewritten in the barred frame as
\begin{equation}
 _V(\vec{x}|\hat{X}_i|\vec{x})_V = \bar{R}_{ij}^{-1}~ _V(\vec{x}|\hat{\bar{X}}_j|\vec{x})_V  \label{unbarred}
\end{equation}
with $\bar{R}\in SO(3)$ being the rotation matrix (\ref{rotmatrix}). Now, we have
\begin{equation}
 _V(\vec{x}|\hat{\bar{X}}_j|\vec{x})_V =  ~_V(\vec{x}|\hat{\bar{x}}_j|\vec{x})_V, ~~~\text{since}  ~~~\hat{\bar{X}}_i |\psi) = |\hat{\bar{x}}_i \psi)
\end{equation}
Then, \begin{equation}
       _V(\vec{x}|\hat{\bar{X}}_j|\vec{x})_V = \frac{\theta^\frac{3}{2}}{(2\pi)^\frac{3}{2}} \int\int d^3p~ d^3p' e^{-\frac{\theta}{4}(\vec{p}^2+\vec{p}'^2)}~e^{-i(\vec{p}-\vec{p}').\vec{x}} (\vec{p}'|\hat{\bar{x}}_j|\vec{p})
      \end{equation}
      
and \begin{eqnarray}
    \nonumber  (\vec{p}'|\hat{\bar{X}}_\alpha|\vec{p})&=& -i\delta(p_3-p'_3)\delta(p_\beta-p'_\beta)e^{-\frac{\theta}{4}(p_\alpha-p'_\alpha)^2 +\frac{i}{2}\theta\epsilon_{\alpha\beta}(p_\alpha-p'_\alpha)p_\beta}\frac{\partial}{\partial p_\alpha}\delta(p_\alpha-p'_\alpha),~~~\alpha,\beta=1,2.\\
   (\vec{p}'|\hat{\bar{X}}_3|\vec{p}) &=& -i\delta(p_1-p'_1)\delta(p_2-p'_2)\frac{\partial}{\partial p_3}\delta(p_3-p'_3)
    \end{eqnarray}
With these, we have
\begin{equation}
 _V(\vec{x}|\hat{\bar{X}}_j|\vec{x})_V = \bar{x}_j
\end{equation}
So, returning to the unbarred fiducial frame, we get from (\ref{unbarred})
\begin{equation}
 _V(\vec{x}|\hat{X}_i|\vec{x})_V = \bar{R}_{ij}^{-1}~ \bar{x}_j = x_i
\end{equation}
and \begin{equation}
      _V(\vec{x}|\hat{P}_i|\vec{x})_V = \frac{\theta^\frac{3}{2}}{(2\pi)^\frac{3}{2}} \int d^3p ~p_i~e^{-\frac{\theta}{2}\vec{p}^2} = 0.
    \end{equation}
Having solved ~ $\langle \hat{Z}_\mu\rangle$~($\mu=1,2,...6$) we now have to calculate the expectation values of the composite 
$ \hat{Z}_\mu\hat{Z}_\nu$ and then its symmetrized and anti-symmetrized expectation values to obtain the Variance matrix and the symplectic matrix.

For this, let us calculate the expectation value of a composite $\{\hat{X}_i,\hat{X}_j\}$. Note that, by itself ~$\hat{X}_i\hat{X}_j$~ doesn't transform as a second rank tensor under $SO(3)$, as was shown in \cite{b6}. Indeed, by defining ~ $\hat{x}^R_i=R_{ij}\hat{x}_j$ ~ ($R \in SO(3)$) one gets for the rotated composite operator 
\begin{equation}
(\hat{x}_i\hat{x}_j)^R = m[\Delta_{\theta}(R)(\hat{x}_i\otimes \hat{x}_j)] = \hat{x}_i^R\hat{x}_j^R + \frac{i}{2}\theta_{ij} - \frac{i}{2}R_{ik}\theta_{kl}(R^T)_{lj}
\end{equation}
through the action of deformed co-product 
\begin{equation}
\Delta_{\theta}(R)=F\Delta_0(R)F^{-1} \label{deformed co product}
\end{equation} 
where 
\begin{equation}
\Delta_0(R)=R\otimes R \label{undeformed product}
\end{equation} 
 is the undeformed co-product and
 \begin{equation}
F=e^{\frac{i}{2}\theta_{ij}\hat{P}_i\otimes\hat{P}_j} \label{twist}
\end{equation}
is the abelian drinfeld twist. Nevertheless its symmetric part (anti-commutator) transforms as a tensor, whereas its antisymmetric part (commutator) transforms as an invariant $SO(3)$ scalar, as was observed earlier \cite{b6}.
\begin{equation}
(\hat{x}_i\hat{x}_j)^R +  (\hat{x}_j\hat{x}_i)^R = \hat{x}_i^R\hat{x}_j^R+\hat{x}_j^R\hat{x}_i^R , ~~~~~\text{while}~~~~~(\hat{x}_i\hat{x}_j)^R -  (\hat{x}_j\hat{x}_i)^R = i\theta_{ij}
\end{equation}
Under the rotation $\bar{R}$, we have
\begin{equation}
_V(\vec{x}|\{\hat{X}_i,\hat{X}_j\}|\vec{x})_V= \bar{R}^{-1}_{im}\bar{R}^{-1}_{jn}~_V(\vec{x}|\{\hat{\bar{X}}_m,\hat{\bar{X}}_n\}|\vec{x})_V
\end{equation}
This gives after a straightforward computation 
\begin{equation}
 _V(\vec{x}|\hat{X}_i\hat{X}_j|\vec{x})_V= x_ix_j+\frac{\theta}{2}\delta_{ij}-\frac{\theta_i\theta_j}{4\theta}+\frac{i}{2}\theta_{ij}.
\end{equation}
so that upon symmetrization and anti-symmetrization, one gets \begin{equation}
       \frac{1}{2}\langle\{\hat{X}_i,\hat{X}_j\}\rangle = x_ix_j+\frac{\theta}{2}\delta_{ij}-\frac{\theta_i\theta_j}{4\theta}.
      \end{equation}
and  \begin{equation}
       \frac{1}{2}\langle [\hat{X}_i,\hat{X}_j] \rangle= \frac{i}{2}\theta_{ij}
     \end{equation}respectively.\\
     
With this, we get the first block ~$V_{XX}$~
\begin{equation}
 \frac{1}{2} \langle\{\hat{X}_i,\hat{X}_j\}\rangle -\langle\hat{X}_i\rangle\langle\hat{X}_j\rangle = \begin{bmatrix}
                  \frac{\theta}{2}-\frac{\theta_1^2}{4\theta}&-\frac{\theta_1\theta_2}{4\theta}&-\frac{\theta_1\theta_3}{4\theta}\\
                  -\frac{\theta_1\theta_2}{4\theta}&\frac{\theta}{2}-\frac{\theta_2^2}{4\theta}&-\frac{\theta_2\theta_3}{4\theta}\\
                  -\frac{\theta_1\theta_3}{4\theta}&-\frac{\theta_2\theta_3}{4\theta}&\frac{\theta}{2}-\frac{\theta_3^2}{4\theta}\\
                 \end{bmatrix}
\end{equation}
of the complete non-commutative Variance matrix
\begin{eqnarray}
V^{\theta} &=&\begin{bmatrix}
      V_{XX}&V_{XP}\\
      V_{PX}&V_{PP}
     \end{bmatrix}\\
&=& \begin{bmatrix}
     \frac{1}{2}\langle\{\hat{X}_i,\hat{X}_j\}\rangle-\langle\hat{X}_i\rangle\langle\hat{X}_j\rangle&\frac{1}{2}\langle\{\hat{X}_i,\hat{P}_j\}\rangle-\langle\hat{X}_i\rangle\langle\hat{P}_j\rangle\\
     \frac{1}{2}\langle\{\hat{P}_i,\hat{X}_j\}\rangle-\langle\hat{P}_i\rangle\langle\hat{X}_j\rangle& \frac{1}{2}\langle\{\hat{P}_i,\hat{P}_j\}\rangle-\langle\hat{P}_i\rangle\langle\hat{P}_j\rangle\\
    \end{bmatrix} \label{var}
\end{eqnarray}
where the expectation values are taken in the Voros states.

Let us now calculate the remaining matrix elements of the $V_{XP}$, $V_{PX}$ and $V_{PP}$ parts of $V$. To begin with, let us consider 
\begin{equation}
 _V(\vec{x}|\{\hat{X}_i,\hat{P}_j\}|\vec{x})_V= \bar{R}^{-1}_{im}\bar{R}^{-1}_{jn}~_V(\vec{x}|\{\hat{\bar{X}}_m,\hat{\bar{P}}_n\}|\vec{x})_V
\end{equation}
Again relating one of the terms in the barred frame as,
\begin{equation}
 _V(\vec{x}|\hat{\bar{X}}_m\hat{\bar{P}}_n|\vec{x})_V= \frac{\theta^\frac{3}{2}}{(2\pi)^\frac{3}{2}} \int\int d^3\bar{p}~ d^3\bar{p}' e^{-\frac{\theta}{4}(\vec{\bar{p}}^2+\vec{\bar{p}}'^2)}~e^{-i(\vec{\bar{p}}-\vec{\bar{p}}').\vec{\bar{x}}}\bar{p}_n (\vec{\bar{p}}'|\hat{\bar{x}}_m|\vec{\bar{p}})
\end{equation} We get for the matrix 
\begin{equation}
\frac{1}{2}[ _V(\vec{x}|\{\hat{\bar{X}}_m,\hat{\bar{P}}_n\}|\vec{x})_V]=\begin{bmatrix}
                                                            0&-\frac{1}{2}&0\\
                                                            \frac{1}{2}&0&0\\
                                                            0&0&0\\
                                                           \end{bmatrix}
\end{equation}

Using the expression of $\bar{R}_{ij}$ in (\ref{rotmatrix}), and the fact that ~$\langle\hat{P}_i\rangle=0$~ we get 
\begin{equation}
V_{XP}= \frac{1}{2} \langle\{\hat{X}_i,\hat{P}_j\}\rangle-\langle\hat{X}_i\rangle\langle\hat{P}_j\rangle=\begin{bmatrix}
                                                            0&-\frac{\theta_3}{2\theta}&\frac{\theta_2}{2\theta}\\
                                                            \frac{\theta_3}{2\theta}&0&-\frac{\theta_1}{2\theta}\\
                                                            -\frac{\theta_2}{2\theta}&\frac{\theta_1}{2\theta}&0\\
                                                           \end{bmatrix}=-V_{PX}
\end{equation}
Finally, for ~$V_{PP}$~ we get
\begin{equation}
V_{PP}=\frac{1}{2} \langle\{\hat{P}_i,\hat{P}_j\}\rangle-\langle\hat{P}_i\rangle\langle\hat{P}_j\rangle= \frac{1}{\theta}\delta_{ij}
\end{equation}
Thus, the complete Variance matrix (\ref{var}) is
\begin{equation}
 V^{\theta}=\begin{bmatrix}
                  \frac{\theta}{2}-\frac{\theta_1^2}{4\theta}&-\frac{\theta_1\theta_2}{4\theta}&-\frac{\theta_1\theta_3}{4\theta}& 0&-\frac{\theta_3}{2\theta}&\frac{\theta_2}{2\theta}\\
                  -\frac{\theta_1\theta_2}{4\theta}&\frac{\theta}{2}-\frac{\theta_2^2}{4\theta}&-\frac{\theta_2\theta_3}{4\theta}& \frac{\theta_3}{2\theta}&0&-\frac{\theta_1}{2\theta}\\
                  -\frac{\theta_1\theta_3}{4\theta}&-\frac{\theta_2\theta_3}{4\theta}&\frac{\theta}{2}-\frac{\theta_3^2}{4\theta}&-\frac{\theta_2}{2\theta}&\frac{\theta_1}{2\theta}&0\\
                   0&\frac{\theta_3}{2\theta}&-\frac{\theta_2}{2\theta}&\frac{1}{\theta}&0&0\\
                  -\frac{\theta_3}{2\theta}&0&\frac{\theta_1}{2\theta}&0&\frac{1}{\theta}&0\\
                   \frac{\theta_2}{2\theta}&-\frac{\theta_1}{2\theta}&0&0&0&\frac{1}{\theta}\\
   \end{bmatrix} \label{noncommvar}
\end{equation}
and the non-commutative symplectic matrix $\Omega^{\theta}$ is
\begin{eqnarray}
 \Omega^{\theta}&=&-i\begin{bmatrix}
          \frac{1}{2}\langle[\hat{X}_i,\hat{X}_j]\rangle&\frac{1}{2}\langle[\hat{X}_i,\hat{P}_j]\rangle\\
          \frac{1}{2}\langle[\hat{P}_i,\hat{X}_j]\rangle&\frac{1}{2}\langle[\hat{P}_i,\hat{P}_j]\rangle\\
         \end{bmatrix}\\
         &=&\frac{1}{2}\begin{bmatrix}
                        0&\theta_3&-\theta_2&1&0&0\\
                        -\theta_3&0&\theta_1&0&1&0\\
                        \theta_2&-\theta_1&0&0&0&1\\
                        -1&0&0&0&0&0\\
                        0&-1&0&0&0&0\\
                        0&0&-1&0&0&0\\
                       \end{bmatrix}
\end{eqnarray}
The corresponding commutative Variance matrix $V^0$ and the symplectic matrix $\Omega^0$ can be obtained from the above respective non-commutative matrices by linear transformations \cite{b12} 
\begin{equation}
 V^0 =MV^{\theta}M^T ~~~~\text{and}~~~~\Omega^0=M\Omega^{\theta} M^T.
\end{equation}
where the matrix $M$ is the transformation matrix which relates the commutative position coordinates to the non-commutative position coordinates as
\begin{equation}
 \hat{X}^c_i= M_{ij}\hat{X}_j ~~;~~ [\hat{X}^c_i , \hat{X}^c_j]=0
\end{equation}
As we have (\ref{transfm.}) \cite{b6}
\begin{equation}
 \hat{X}^c_i= \hat{X}_i+\frac{\theta_{ij}}{2}\hat{P}_j, 
\end{equation}
the matrix $M$ is given by
\begin{equation}
 M= \begin{bmatrix}
     1&0&0&0&\frac{\theta_3}{2}&-\frac{\theta_2}{2}\\
     0&1&0&-\frac{\theta_3}{2}&0&\frac{\theta_1}{2}\\
     0&0&1&\frac{\theta_2}{2}&-\frac{\theta_1}{2}&0\\
     0&0&0&1&0&0\\
     0&0&0&0&1&0\\
     0&0&0&0&0&1\\
    \end{bmatrix}
\end{equation}
With this, the commutative variance matrix and the commutative symplectic matrix is found to be
\begin{equation}
 V^0=\begin{bmatrix}
      \frac{\theta}{4}&0&0&0&0&0\\
      0&\frac{\theta}{4}&0&0&0&0\\
      0&0&\frac{\theta}{4}&0&0&0\\
      0&0&0&\frac{1}{\theta}&0&0\\
      0&0&0&0&\frac{1}{\theta}&0\\
      0&0&0&0&0&\frac{1}{\theta}\\
     \end{bmatrix} \label{comsymmat}
\end{equation}
and 
\begin{equation}
\Omega^0=\frac{1}{2i}\langle [{\hat{Z}^0}_\mu,{\hat{Z}^0}_\nu]\rangle=\frac{1}{2}\begin{bmatrix}
                            0&0&0&1&0&0\\
                            0&0&0&0&1&0\\
                            0&0&0&0&0&1\\
                            -1&0&0&0&0&0\\
                            0&-1&0&0&0&0\\
                            0&0&-1&0&0&0\\
                           \end{bmatrix}
\end{equation}
with $ {\hat{Z}^0}_i={\hat{X}^c}_i ~ ~, ~ ~{\hat{Z}^0}_{i+3}=\hat{P}_i~~(i=1,2,3)$. Then calculating the symplectic eigenvalues of $V^0$ i.e. the ordinary eigenvalues of $|2i\Omega^0V^0|$ \cite{b8}, we get the eigenvalues as 
$\frac{1}{2},\frac{1}{2},\frac{1}{2},\frac{1}{2},\frac{1}{2},\frac{1}{2}$ i.e. 6-fold degenerate. These three pairs of symplectic eigenvalues, each of the form of $(\frac{1}{2},\frac{1}{2})$ can be obtained more simply from the corresponding single mode $V^0$ of the form 
$(\frac{\theta}{4},\frac{1}{\theta})$ (\ref{comsymmat}), which occurs symmetrically in all the directions $x,y,z$, by a simple canonical transformation $(x,p)\rightarrow (\lambda x , \frac{1}{\lambda} p)$ for suitable 
$\lambda \neq 0$. In any case it is simple to see that both $V^{\theta}$ (\ref{noncommvar}) and $V^0$ (\ref{comsymmat}) satisfy the saturation condition (\ref{satcon}) (since $\det M = 1$)
\begin{equation}
\det V^{\theta}= \det V^0= \frac{1}{4^3}
\end{equation}
This indicates that the Voros basis represents a maximally localized state in phase space. But note that the Voros basis (\ref{voros}), can be factorized by going to the barred frame as 
\begin{equation}
|\vec{x})_V= \Big(\frac{\theta}{2\pi} \int d^2 \bar{p} ~ e^{-\frac{\theta}{4}({\bar{p}_1}^2 + {\bar{p}_2}^2)} e^{i\bar{p}_{\alpha}(\hat{\bar{x}}_{\alpha} - \bar{x}_{\alpha})}\Big)\Big(\sqrt{\frac{\theta}{2\pi}}\int d\bar{p}_3 e^{-\frac{\theta}{4} {\bar{p}_3}^2} e^{i\bar{p}_3(\hat{\bar{x}}_3 -\bar{x}_3)}\Big)
\end{equation}
where the first factor represents the 2D Voros basis (\ref{vorosbasis}) and the second factor $\int d\bar{p}_3 e^{-\frac{\theta}{4} {\bar{p}_3}^2} e^{i\bar{p}_3(\hat{\bar{x}}_3 -\vec{x}_3)} \sim e^{-\frac{1}{\theta}(\hat{\bar{x}}_3 -\bar{x}_3)^2}$, representing a one-dimensional Gaussian state centered at $\bar{x}_3$ with a spread $\Delta \bar{x}_3 \sim \sqrt{\theta}$. Clearly, this $\Delta \bar{x}_3 $ can be made as small as we like by a suitable scaling factor and scaling up $\Delta \bar{p}_3$ appropriately to preserve the saturation condition (obviously this is a non-Voros state). The generalized points, i.e. pure states of the involutive $C^*-$algebra, which is just $\mathcal{H}_q$ in this case can be described by using the barred variables by the density matrices $|z,\bar{x}_3\rangle \langle z,\bar{x}_3 | \in \mathcal{H}_q$. Correspondingly, the pure states in the fiducial frame is obtained by applying a unitary transformation: $U(\bar{R})|z,\bar{x}_3\rangle \langle z,\bar{x}_3 | U^\ddagger(\bar{R}) $. 
This non-Voros state will be useful if we are interested in computing Connes' spectral distance between a pair of such  \textquotedblleft points \textquotedblright~ i.e. pure states (see for example \cite{b16} and references there-in). However, since our interest is to compute the thermal correlation function between a pair of identical particles in Voros states, we shall not be concerned with such states in the rest of the paper. Thus even if $\Delta \bar{x}_3$ is squeezed to the extreme such that $\Delta \bar{x}_3 = 0$  we can 
still have $\Delta \bar{X}_1 \Delta \bar{X}_2 \geq \frac{\theta}{2}$ for such a non-Voros state. More generally, we can write in this case
\begin{equation}
\Delta \bar{X}_1 \Delta \bar{X}_2 + \Delta \bar{X}_2 \Delta \bar{X}_3 + \Delta \bar{X}_3 \Delta \bar{X}_1 \geq \frac{\theta}{2}
\end{equation}
It is therefore quite interesting to see the form of analogous inequality in the original fiducial frame for the Voros basis.

 Using $\Delta X_i = \sqrt{\frac{\theta}{2}-\frac{\theta_i^2}{4\theta}}$ from (\ref{noncommvar}), we have 
 \begin{equation}
 \Delta \hat{X}_1 \Delta \hat{X}_2 + \Delta \hat{X}_2 \Delta \hat{X}_3 + \Delta \hat{X}_1 \Delta \hat{X}_3 = \frac{1}{4\theta} \left[\sqrt{(2\theta^2-\theta_1^2) (2\theta^2-\theta_2^2)}+\sqrt{(2\theta^2-\theta_2^2)(2\theta^2-\theta_3^2)} +\sqrt{(2\theta^2-\theta_1^2)(2\theta^2-\theta_3^2)}\right]  \label{spaceuncertainty}
 \end{equation}
where the vector $\vec{\theta}$ points in arbitrary direction. We can see that the expression (\ref{spaceuncertainty}) attains its minimum value $\frac{\theta}{2}(1+\sqrt{2})$ when the vector $\vec{\theta}$ points in one of the three axes (e.g. $\theta=\theta_3;~~\theta_1=\theta_2=0$). So we have the following condition
\begin{equation}
\Delta \hat{X}_1 \Delta \hat{X}_2 + \Delta \hat{X}_2 \Delta \hat{X}_3 + \Delta \hat{X}_1 \Delta \hat{X}_3 \geq \frac{\theta}{2}(1+\sqrt{2}).
\end{equation}

\section{Two particle formalism}
\label{sec:Two particle formalism}
Let us extend this formulation to two particle systems. In commutative quantum mechanics the way we proceed is to think of the two particle system in terms of wave-functions defined over $R^6$ and therefore to think of classical configuration space as a tensor product $R^3\otimes R^3$.  One may therefore be tempted to take the same approach in non-commutative quantum mechanics and to introduce the non-commutative $3D$ configuration space for a two particle system as a tensor product of two single particle configuration spaces, i.e., ${\mathcal{H}_c}^{(2)}=  \mathcal{H}_c\otimes  \mathcal{H}_c.$ and the quantum Hilbert space as the space of operators generated by the $\hat{\vec{x}}_1$ and $\hat{\vec{x}}_2$, with the subscripts referring to the non-commutative coordinates of the two particles, i.e., the elements of the quantum Hilbert space are operators $\psi(\hat{\vec{x}}_1, \hat{\vec{x}}_2)$. This is essentially the approach adopted in \cite{b9} where a Moyal-like star product between the functions $\varPsi(
\vec{x}_1,\vec{x}_2)$ and $ \varPhi(\vec{x}_1,\vec{x}_2)$ was introduced as 
\begin{equation}
 (\varPsi\star\varPhi)(\vec{x}_1,\vec{x}_2)=\varPsi(\vec{x}_1,\vec{x}_2)e^{\frac{i}{2}\theta_{ij}(\frac{\overleftarrow{\partial}}{\partial x_{1i}}+\frac{\overleftarrow{\partial}}{\partial x_{2i}})(\frac{\overrightarrow{\partial}}{\partial x_{1j}}+\frac{\overrightarrow{\partial}}{\partial x_{2j}})}\varPhi(\vec{x}_1,\vec{x}_2) \label{A}
\end{equation}
which yields the following commutation relations
\begin{equation}
 [\hat{x}_{\alpha i}, \hat{x}_{\beta j}]= i\theta_{ij} \label{commuationrelation}
\end{equation}
where the subscript $\alpha,\beta$ are the particle labels. These types of commutation relations are also obtained in the approach of braided twisted symmetry \cite{b10}.

This way of thinking is, however, very misleading and, indeed, inappropriate in the context of the operator algebra formulation of non-commutative quantum mechanics outlined in section \ref{section2},  as we now proceed to argue.  The more appropriate way to think about a two particle commutative system is to think of it as two sets of $3D$ labels in the same $3D$ configuration space or, more appropriately, in the language of relativity as the coordinates of two \textquotedblleft events" in $3D$.    In non-commutative quantum mechanics the notion of coordinates does not exists in classical configuration space and should arise through the state $|\psi(\hat{\vec{x}}))$ or operator $\psi(\hat{\vec{x}})$, which acts  on $\mathcal{H}_c$, and describes the quantum state of the system.  The real issue, therefore, is to identify the states in quantum Hilbert space that would describe a two particle state with particles localised at points $\vec{x}_1$ and $\vec{x}_2$.  The answer to this question is actually quite natural and simple. 
 To start, let us first consider one particle and, for demonstrative purposes, in $2D$.  We want to find the state in quantum Hilbert space describing a particle localised at $z=(x_1+ix_2)/\sqrt{2}$.  The state with this property is the one given by $|z,n)=|z\rangle\langle n|$.  As the action of the position operators are defined by left action,  it is an eigenstate of  $X_1+iX_2$ and subsequently a minimum uncertainty state in classical configuration space.  The right hand sector describes another property of the system.  A more detailed discussion of this interpretation can be found in \cite{b19}.  This interpretation is also borne out by calculating the position representation of this state, e.g., 
\begin{equation}
\label{posstate}
(w|z,n)=e^{-\frac{1}{2}|z-w|^2} e^{\frac{1}{2}(\bar w z-w\bar z)}\langle n|w\rangle.
\end{equation}
setting $\xi=w-z$, this turns into
\begin{equation}
\label{posstate1}
(w|z,n)=e^{-\frac{1}{2}|\xi|^2} e^{\frac{1}{2}(z \bar\xi-\bar z\xi)}\langle n|z+\xi\rangle,
\end{equation}
clearly demonstrating that the state is localised at $z$ with non-local corrections deriving from an expansion in $\xi$.   Restoring dimensions, $\xi$ is of the order $\sqrt{\theta}$, demonstrating that the non-local corrections are of the order of the length scale set by the non-commutative parameter.  Keeping in mind that $|z,n)$ is an operator on $\mathcal{H}_c$, and therefore an element of the algebra generated by the $\hat{x}_i$, it can be written in the form $|z,n)=|\psi_{z,n}(\hat{\vec{x}}))\equiv |\psi_{\vec{x},n}(\hat{\vec{x}}))$ (note that $\vec{x}$ is a label and $\hat{\vec{x}}$ are operators).  

The two particle state must now also be an operator acting on the same configuration space $\mathcal{H}_c$ and therefore be in the algebra generated by the $\hat{x}_i$, but it must also carry the indices of two particles.  The most obvious, and probably only, construction of such two particle states is:
\begin{equation}
|\vec{x}_1,n_1;\vec{x}_2,n_2)=|\psi_{\vec{x}_1,n_1}(\hat{\vec{x}})\psi_{\vec{x}_2,n_2}(\hat{\vec{x}}))\equiv |(\psi_{\vec{x}_1,n_1}\psi_{\vec{x}_2,n_2})(\hat{\vec{x}}))\equiv |(\psi_1\psi_2)(\hat{\vec{x}}))
\end{equation}

This is a state in quantum Hilbert space representing two particles localised at points $\vec{x}_1$ and $\vec{x}_2$ as can also be borne out by computing the position presentation of this state.  Indeed, the wave function of the state $|(\psi_1\psi_2)(\hat{\vec{x}}))$ is obtained by taking the overlap in M/V basis to yield a M/V star product composed wave function
\begin{equation}
 _{M/V}(\vec{x}\mid \psi_1(\hat{\vec{x}})\psi_2(\hat{\vec{x}})) = \lambda_{M/V} {~}_{M/V}(\vec{x}|\psi_1(\hat{\vec{x}})){~}\star_{M/V} {~}_{M/V}(\vec{x}|\psi_2(\hat{\vec{x}})) 
\end{equation}
where $\lambda_M = 2\pi\theta^\frac{3}{4}$,  $\lambda_V = 1$ and the states $(\vec{x}|\psi_i(\hat{\vec{x}}))$, $i=1,2$ are of the form (\ref{posstate}) or (\ref{posstate1}).  The two particle nature of this construction can also be made more explicit by  interpreting it as a map $m:\mathcal{H}_q\otimes \mathcal{H}_q\rightarrow \mathcal{H}_q$:
\begin{equation}
\label{2particlestates}
m[|\psi_1(\hat{\vec{x}}))\otimes |\psi_2(\hat{\vec{x}}))]= |(\psi_1\psi_2)(\hat{\vec{x}})).
\end{equation}
This also brings us in contact with the philosophy of \cite{b5} and the implementation of twisting adopted in \cite{b14}.

We can now address the question of the transformational property of a generic two-particle state (\ref{2particlestates}) under an infinitesimal SO(3) rotation. This is determined by using a mathematical consistency condition following from an identity in Hopf algebra. This is given in this context as
\begin{equation}
 \hat{J}_i\left[m\left(|\psi_1(\hat{\vec{x}}))\otimes|\psi_2(\hat{\vec{x}}))\right)\right] = m\left[\Delta_\theta(\hat{J}_i)\left(|\psi_1(\hat{\vec{x}}))\otimes|\psi_2(\hat{\vec{x}}))\right)\right] 
\end{equation}

As it has been already obtained in \cite{b6} the co product of the angular momentum operators $\hat{J_i}$ get deformed in the non-commutative $3D$ Quantum Hilbert space. So the action of this deformed co-product on a generic two particle state can be obtained as
\begin{eqnarray} \label{102}
 \Delta_\theta(\hat{J}_i)(\psi_1(\hat{\vec{x}})\otimes\psi_2(\hat{\vec{x}})) = \hat{J}_i\psi_1(\hat{\vec{x}})\otimes\psi_2(\hat{\vec{x}}) + \psi_1(\hat{\vec{x}})\otimes\hat{J}_i\psi_2(\hat{\vec{x}}) \\ \nonumber
 + \frac{1}{2}[\hat{P}_i\psi_1(\hat{\vec{x}})\otimes(\vec{\theta}.\vec{P})\psi_2(\hat{\vec{x}}) - (\vec{\theta}.\vec{P})\psi_1(\hat{\vec{x}})\otimes\hat{P}_i\psi_2(\hat{\vec{x}})] 
\end{eqnarray}
so that, using the twist (\ref{twist})
\begin{equation}
\Delta_{\theta}(\hat{J}_i) = \hat{J}_i \otimes I + I \otimes \hat{J}_i + \frac{1}{2}\left[\hat{P}_i \otimes (\vec{\theta} \cdot \vec{P})  - (\vec{\theta}\cdot \vec{P}) \otimes \hat{P}_i \right]  = F  \Delta_0 (\hat{J}_i) F^{-1}
\end{equation}
Let us now introduce an exchange operation $\textquoteleft\Sigma$' on the quantum Hilbert space ${\mathcal{H}_q}^{(2)}$ such that
\begin{equation}
 \Sigma : A \otimes B \rightarrow B \otimes A \label{exchangeformalism} 
\end{equation}
we find 
\begin{equation}
 \Sigma[\Delta_\theta(\hat{J}_i)(\psi_1(\hat{\vec{x}})\otimes\psi_2(\hat{\vec{x}}))]\neq \Delta_\theta(\hat{J}_i)[\Sigma(\psi_1(\hat{\vec{x}})\otimes\psi_2(\hat{\vec{x}})]
\end{equation}
This implies that the under a transformation like rotation the statistics of the physical state can get altered; a pure bosonic/fermionic state, obtained by projecting into symmetric/antisymmetric subspace by the projector \\
 $P^{\pm}=\frac{1}{2}(I\pm\Sigma)$ will yield an admixture of bosonic/fermionic states under rotation. But this cannot be allowed as implied by the super selection rules which says that a system of fermions or bosons should remain as the one under any transformation. And for this, the exchange operation should commute with the deformed co-product. Hence, the exchange operator $\textquoteleft\Sigma$' should also get deformed as 
\begin{equation}
 \Sigma_\theta = F~\Sigma ~ F^{-1} \label{tpf25}
\end{equation}
such that    ~~~ $[\Sigma_\theta, \Delta_\theta] = 0 $.

Corresponding to this deformed exchange operator, the deformed projection operator can be constructed as 
\begin{equation}
 P_\theta^{\pm} = \frac{1}{2}(I \pm \Sigma_\theta )\label{2projection}
\end{equation}
Since
 $ F = e^{{\frac{i}{2}\theta_{ij}\hat{P}_i\otimes\hat{P}_j}} $, we can easily check that $\Sigma F^{-1} = F\Sigma$ so that
\begin{equation}
 \Sigma_\theta = F^2 \Sigma = e^{i\theta_{ij}\hat{P}_i\otimes\hat{P}_j}\Sigma \label{tpf27}
\end{equation}
We then obtain
\begin{equation}
 P_\theta^{\pm}(\psi_1(\hat{\vec{x}})\otimes \psi_2(\hat{\vec{x}})) = \frac{1}{2}[\psi_1(\hat{\vec{x}})\otimes\psi_2(\hat{\vec{x}}) \pm e^{i{\theta_i}_j\hat{P}_i\otimes\hat{P}_j} (\psi_2(\hat{\vec{x}})\otimes\psi_1(\hat{\vec{x}}))]\label{projection} 
\end{equation}
Here, $P_\theta^{\pm}$ is referred as the twisted symmetric(+)/ antisymmetric(-) projection operator which give the twisted symmetric/ antisymmetric states corresponding to the twisted bosons/ fermions system.\\

\section{Many-particle states}

The construction of N-particle states proceed in complete analogy with the two particle states.  The states in quantum Hilbert space representing N-particle states are of the form 
\begin{equation}
 |\psi_{\vec{x}_1,n_1}(\hat{\vec{x}})\psi_{\vec{x}_2,n_2}(\hat{\vec{x}})\ldots\psi_{\vec{x}_N,n_N}(\hat{\vec{x}}))\equiv |(\psi_{\vec{x}_1,n_1}\psi_{\vec{x}_2,n_2}\ldots\psi_{\vec{x}_N,n_N})(\hat{\vec{x}}))\equiv |(\psi_1\psi_2\ldots\psi_N)(\hat{\vec{x}})).
\end{equation}
This can again be interpreted as a map:
\begin{equation}
m\left [\prod_{i=1}^N\otimes |\psi_i(\hat{\vec{x}}))\right]=|(\psi_1\psi_2\ldots\psi_N)(\hat{\vec{x}})).
\end{equation}

We have already mentioned that for two-particles system, the deformed co-product gives rise to the deformed permutation symmetry so that a twisted symmetric or antisymmetric two particle state is given by (\ref{projection})
\begin{equation}
 \Psi(\hat{x}_1,\hat{x}_2)_{\pm\theta}\equiv|\Psi)_{\pm\theta}= |\psi_1(\hat{x}), \psi_2(\hat{x}))_{\pm\theta} =P^{\pm}_\theta(\psi_1(\hat{x})\otimes\psi_2(\hat{x})).  
 \end{equation}
Here, $\textquoteleft+\theta$' corresponds to the twisted symmetric state and $\textquoteleft-\theta$' corresponds to the twisted antisymmetric state.
\subsection{Twisted symmetric/antisymmetric momentum basis}
The corresponding twisted symmetric/antisymmetric momentum eigenstates can be written more compactly as 
\begin{equation}
 |\vec{p}_1,\vec{p}_2)_{\theta} = \frac{1}{2}[|\vec{p}_1)\otimes|\vec{p}_2)+ \eta e^{i\theta_{ij}p_{2i}p_{1j}}|\vec{p}_2)\otimes|\vec{p}_1)] \label{twisted2basis}
 \end{equation}
 where $\eta = +1$ corresponds to twisted symmetric states and $\eta = -1$ corresponds to twisted antisymmetric states.
 
These states do not satisfy the orthonormality conditions but do satisfy the completeness relation:

\begin{equation}
 _{\theta}(\vec{p}'_1, \vec{p}'_2|\vec{p}_1, \vec{p}_2)_{\theta} = \frac{1}{2}[\delta^3(\vec{p}'_1-\vec{p}_1)\delta^3(\vec{p}'_2-\vec{p}_2) +\eta e^{i\theta_{ij}p_{2i}p_{1j}}\delta^3(\vec{p}'_1-\vec{p}_2)\delta^3(\vec{p}'_2-\vec{p}_1)] \label{twistedorthonormality}
 \end{equation}

and 
\begin{eqnarray}
 \int d^3 p_1 d^3 p_2 |\vec{p}_1, \vec{p}_2)_{\theta} ~_{\theta}(\vec{p}_1, \vec{p}_2| = \mathbf{1}_q
\end{eqnarray}

We can extend this to N-particle system so that the twisted N-particle symmetric and antisymmetric states \cite{b14} can be constructed as
\begin{equation}
 |\psi_1, \psi_2,....,\psi_N)_{\theta} = P^{N}_\theta(\psi_1\otimes\psi_2\otimes...\otimes\psi_N)
\end{equation}
We just need to find the form of the deformed projection operator $P^N$ for N-particle system. For 2-particle system, it is given by (\ref{2projection}). In order to extend to 3-particle system, first we should note that here we have to encounter with two deformed nearest neighbor exchange operators: $\Sigma_\theta^{12}=\Sigma_\theta\otimes 1$ which exchanges the first and the second slots keeping the third slot fixed in the tensor product of three operators and $\Sigma_\theta^{23}=1\otimes\Sigma_\theta$ which exchanges the second and the third slots keeping the first slot fixed. Thus, we can write the deformed projection operator for 3-particle physical states as
\begin{equation}
 P^{3}_\theta = \frac{1}{3!}[1~+\eta~\{\Sigma^{12}_\theta~+~\Sigma^{23}_\theta\}~+\eta^2~\{\Sigma^{12}_\theta~\Sigma^{23}_\theta~+~\Sigma^{23}_\theta~\Sigma^{12}_\theta\}~+\eta^3~\Sigma^{12}_\theta~\Sigma^{23}_\theta~\Sigma^{12}_\theta]
\end{equation}
where we should note that 
\begin{equation}
 (\Sigma^{12}_\theta)^2 = 1= (\Sigma^{23}_\theta)^2 \label{3first}
\end{equation}
and 
\begin{equation}
 \Sigma^{12}_\theta~\Sigma^{23}_\theta~\Sigma^{12}_\theta=\Sigma^{23}_\theta~\Sigma^{12}_\theta~\Sigma^{23}_\theta \label{3second}
\end{equation}

Then for the N-particle system, there should be (N-1) deformed nearest neighbor exchange operators~ $\Sigma^{n,n+1}_\theta,~~n=1,2....(N-1)$ which exchanges only the entries at the $n$th and $(n+1)$th slots, keeping all the entries at other slots fixed. That is, 
\begin{equation}
 \Sigma_\theta^{n,n+1} = 1\otimes 1\otimes....\otimes\Sigma_\theta\otimes....\otimes 1\otimes 1, ~~~\Sigma_\theta~~ \text{is at the $n$th position}.
\end{equation}

The last two relations for 3-particle system (\ref{3first}, \ref{3second}) can be put in a generalized form for an N-particle system as
\begin{equation}
 (\Sigma^{n,n+1}_\theta)^2 = 1, 
\end{equation}
and
\begin{equation}
\Sigma^{n,n+1}_\theta~\Sigma^{n+1,n+2}_\theta~\Sigma^{n,n+1}_\theta=\Sigma^{n+1,n+2}_\theta~\Sigma^{n,n+1}_\theta~\Sigma^{n+1,n+2}_\theta.
\end{equation}
so that the deformed projection operator for N-particle system is given by 
\begin{eqnarray}
 P^N_\theta &=& \frac{1}{N!}\sum_{n=1}^{N-1}[1~+ \eta~\Sigma^{n,n+1}_\theta~+\eta^2~\Sigma^{n,n+1}_\theta~\Sigma^{n+1,n+2}_\theta~+\eta^3..~\\ \nonumber
 &&......+\eta^N~\Sigma^{n,n+1}_\theta\Sigma^{n+1,n+2}_\theta\Sigma^{n+2,n+3}_\theta.....]
\end{eqnarray}
We should not forget that ~$\Sigma_\theta^{n,n+1} = F^2\Sigma = e^{i\theta^{ij}\hat{P}_i\otimes\hat{P}_j}\Sigma$~ so that in the above expression we have a phase factor for each deformed exchange operator. Thus in the last term of the above equation there are N phase factors for N deformed exchange operators. Needless to say that $\eta^N=\pm$, depending on whether N is even or odd.

In this way, we can define a twisted symmetric/ antisymmetric state corresponding to the twisted bosons/fermions for N-particle system on the quantum Hilbert space $\mathcal{H}^{(3)}_q$. It is obvious that we can get the corresponding twisted symmetric/antisymmetric momentum eigenstates as
\begin{eqnarray}
 |\vec{p}_1, ..,\vec{p}_n,\vec{p}_{n+1}..,\vec{p}_N)_{\theta} = P^{N}_\theta(|\vec{p}_1)\otimes..\otimes|\vec{p}_n)\otimes|\vec{p}_{n+1})\otimes..\otimes|\vec{p}_N)) \\ \nonumber
 = \frac{1}{N!}\sum_{n=1}^{N-1}[|\vec{p}_1)\otimes..\otimes|\vec{p}_n)\otimes|\vec{p}_{n+1})\otimes..\otimes|\vec{p}_N) \\ \nonumber
  + \eta e^{ip_{n+1}\wedge p_n}|\vec{p}_1)\otimes ...\otimes|\vec{p}_{n+1})\otimes|\vec{p}_n)\otimes...\otimes|\vec{p}_N) \\ \nonumber
  +\eta^2 e^{ip_{n+1}\wedge p_n}e^{ip_{n+2}\wedge p_{n+1}}\{|\vec{p}_1)\otimes..\otimes|\vec{p}_{n+2})\otimes|\vec{p}_n)\otimes|\vec{p}_{n+1})\otimes..\otimes|\vec{p}_N) \\ \nonumber
  +~|\vec{p}_1)\otimes..\otimes|\vec{p}_{n+1})\otimes|\vec{p}_{n+2})\otimes|\vec{p}_n)\otimes..\otimes|\vec{p}_N)\}+............................ \\ \nonumber
+\eta^N e^{ip_{n+1}\wedge p_n}e^{ip_{n+2}\wedge p_{n+1}}.........|\vec{p}_N) \otimes....\otimes|\vec{p}_{n+1})\otimes|\vec{p}_{n})\otimes....\otimes|\vec{p}_1)]
\end{eqnarray}
where we have used the wedge $\textquoteleft\wedge$' between the momenta to simply denote the following factor for simplicity, it has no relation with the wedge product of differential geometry.
\begin{equation}
 p\wedge p' = \theta^{ij}p_ip'_j.
\end{equation}
We can obtain the following relation for the twisted symmetric/antisymmetric momentum eigenstates for such N-particle system as
\begin{eqnarray}
_{\theta}(\vec{p}'_1,\vec{p}'_2. ...,\vec{p}'_N|\vec{p}_1,\vec{p}_2 ,..,\vec{p}_N)_{\theta} = \frac{1}{N!}\sum_{n=0}^{(N-1)}[\delta^3(\vec{p}'_1-\vec{p}_1)\delta^3(\vec{p}'_2-\vec{p}_2)...\delta^3(\vec{p}'_N-\vec{p}_N)\\ \nonumber
+\eta e^{ip_{n+1}\wedge p_n}\{\delta^3(\vec{p}'_1-\vec{p}_1).........\delta^3(\vec{p}'_n-\vec{p}_{n+1}).........\delta^3(\vec{p}'_{n+1}-\vec{p}_n).........\delta^3(\vec{p}'_N-\vec{p}_N)\}\\ \nonumber
+\eta^2e^{ip_{n+1}\wedge p_n}e^{ip_{n+2}\wedge p_{n+1}}\{\delta(\vec{p}'_1-\vec{p}_1)........\delta^3(\vec{p}'_n-\vec{p}_{n+2})\delta^3(\vec{p}'_{n+1}-\vec{p}_n)\delta^3(\vec{p}'_{n+2}-\vec{p}_{n+1})\\ \nonumber
.......\delta^3(\vec{p}'_N-\vec{p}_N) + \delta^3(\vec{p}'_1-\vec{p}_1).............\delta^3(\vec{p}'_n-\vec{p}_{n+1})\delta^3(\vec{p}'_{n+1}-\vec{p}_{n+2}) \delta^3(\vec{p}'_{n+2}-\vec{p}_n)\\ \nonumber
................\delta^3(\vec{p}'_N-\vec{p}_N)\} + ..............................+\eta^Ne^{ip_{n+1}\wedge p_n}e^{ip_{n+2}\wedge p_{n+1}}..................\\ \nonumber
\times\delta^3(\vec{p}'_1-\vec{p}_N)  \delta^3(\vec{p}'_2-\vec{p}_{N-1})............................................\delta^3(\vec{p}'_{N-1}-\vec{p}_2)\delta^3(\vec{p}'_N-\vec{p}_1)].\\ \nonumber
\end{eqnarray}
Clearly, this demonstrates the non-orthogonality between these pair of states in the N-particle sector, because of the presence of relative phase factors. However, the resolution of identity is satisfied and the corresponding completeness relation is given as
\begin{equation}
 \int d^3p_1d^3p_2....d^3p_N|\vec{p}_1,\vec{p}_2 ...,\vec{p}_N)_{\theta}~_{\theta}(\vec{p}_1,\vec{p}_2 ....,\vec{p}_N|= \mathbf{1}^{N}_q. \label{twistedcommcompleteness}
\end{equation}
Since these twisted symmetric/anti-symmetric momentum eigenstates are not orthogonal, we search for another basis (in the next subsection) which is  just different from this former basis by a phase factor and check whether it is orthogonal, without upsetting the completeness relation.

\subsection{Quasi-commutative symmetric/antisymmetric momentum basis}
We can now introduce a new basis $|\vec{p}_1,\vec{p}_2))$  (denoted henceforth by a \textquotedblleft double ket \textquotedblright) as
\begin{equation}
 |\vec{p}_1,\vec{p}_2)) = \frac{1}{2}[e^{\frac{i}{2}p_1\wedge p_2}|\vec{p}_1)\otimes |\vec{p}_2) + \eta e^{\frac{i}{2}p_2\wedge p_1}|\vec{p}_2)\otimes |\vec{p}_1)] \label{quasi2basis}
\end{equation} 
This basis will be referred as quasi-commutative basis and is found to be symmetric under the undeformed exchange operation $\Sigma$ (\ref{exchangeformalism}) and it is related to the twisted basis (\ref{twisted2basis}) as
\begin{equation}
 |\vec{p}_1, \vec{p}_2)_{\theta} = e^{-\frac{i}{2}p_1\wedge p_2}|\vec{p}_1,\vec{p}_2))
\end{equation}
We can easily check that this new basis satisfy the usual orthogonality and completeness relations as
\begin{equation}
 ((\vec{p}'_1,\vec{p}'_2|\vec{p}_1, \vec{p}_2)) = \frac{1}{2}[\delta^3(\vec{p}'_1-\vec{p}_1)\delta^3(\vec{p}'_2-\vec{p}_2) +\eta \delta^3(\vec{p}'_1-\vec{p}_2)\delta^3(\vec{p}'_2-\vec{p}_1)]
\end{equation}
and
\begin{equation}
 \int d^3 p_1 d^3 p_2 ~|\vec{p}_1,\vec{p}_2))((\vec{p}_1,\vec{p}_2| = \mathbf{1}_q.
\end{equation}
In the similar way we can define the new basis for 3-particle and so on as
\begin{eqnarray}
 |\vec{p}_1,\vec{p}_1, \vec{p}_3))_\pm &=& \frac{1}{3!}[e^{\frac{i}{2}(p_1\wedge p_2 + p_2\wedge p_3 + p_1\wedge p_3)}|\vec{p}_1)\otimes |\vec{p}_2)\otimes |\vec{p}_3)
 + \eta e^{\frac{i}{2}(p_2\wedge p_1 + p_2\wedge p_3 + p_1\wedge p_3)}|\vec{p}_2)\otimes |\vec{p}_1)\otimes |\vec{p}_3) \\ \nonumber
& + & \eta e^{\frac{i}{2}(p_1\wedge p_2 + p_3\wedge p_2 + p_1\wedge p_3)}|\vec{p}_1)\otimes |\vec{p}_3)\otimes |\vec{p}_2) + \eta^2 e^{\frac{i}{2}(p_1\wedge p_2 + p_3\wedge p_2 + p_3\wedge p_1)}|\vec{p}_3)\otimes |\vec{p}_1)\otimes |\vec{p}_2) \\ \nonumber
& + & \eta^2 e^{\frac{i}{2}(p_2\wedge p_1 + p_2\wedge p_3 + p_3\wedge p_1)}|\vec{p}_2)\otimes |\vec{p}_3)\otimes |\vec{p}_1)+\eta^3 e^{\frac{i}{2}(p_2\wedge p_1 + p_3\wedge p_2 + p_3\wedge p_1)}|\vec{p}_3)\otimes |\vec{p}_2)\otimes |\vec{p}_1)]
\end{eqnarray} where 
\begin{equation}
|\vec{p}_1,\vec{p}_2,\vec{p}_3)_\theta = e^{-\frac{i}{2}(p_1\wedge p_2+p_2\wedge p_3+p_1\wedge p_3)}|\vec{p}_1,\vec{p}_1, \vec{p}_3))_\pm                       
\end{equation}
satisfying the following corresponding orthogonality and completeness relations 
\begin{eqnarray}
 \nonumber_\pm((\vec{p}'_1,\vec{p}'_2,\vec{p}'_3|\vec{p}_1,\vec{p}_2,\vec{p}_3))_\pm &=& \frac{1}{3!}[\delta^3(\vec{p}'_1-\vec{p}_1)\delta^3(\vec{p}'_2-\vec{p}_2)\delta^3(\vec{p}'_3-\vec{p}_3)
 +\eta\delta^3(\vec{p}'_1-\vec{p}_2)\delta^3(\vec{p}'_2-\vec{p}_1)\delta^3(\vec{p}'_3-\vec{p}_3)\\ \nonumber
 &+&\eta\delta^3(\vec{p}'_1-\vec{p}_1)\delta^3(\vec{p}'_2-\vec{p}_3)\delta^3(\vec{p}'_3-\vec{p}_2)
 + \eta^2\delta^3(\vec{p}'_1-\vec{p}_3)\delta^3(\vec{p}'_2-\vec{p}_1)\delta^3(\vec{p}'_3-\vec{p}_2)\\ \nonumber
 &+& \eta^2\delta^3(\vec{p}'_1-\vec{p}_2)\delta^3(\vec{p}'_2-\vec{p}_3)\delta^3(\vec{p}'_3-\vec{p}_1)
+\eta^3\delta^3(\vec{p}'_1-\vec{p}_3)\delta^3(\vec{p}'_2-\vec{p}_2)\delta^3(\vec{p}'_3-\vec{p}_1)]
\end{eqnarray} and
\begin{equation}
 \int d^3 p_1 d^3 p_2 d^3 p_3 ~|\vec{p}_1,\vec{p}_2,\vec{p}_3)) ((\vec{p}_1,\vec{p}_2,\vec{p}_3| = \mathbf{1}_q. \label{quasicomm.}
\end{equation}
In this way, we can write such symmetrized/anti-symmetrized quasi-commutative basis for any arbitrary number of particles which satisfy the usual orthogonality and completeness relations.
\section{Second Quantization}
Let us now enlarge the physical quantum Hilbert space to include states of arbitrary number of particles. We can define a full quantum Hilbert space which is just the direct sum of the spaces with all possible number particle states. That is,
\begin{equation}
\mathcal{H}_Q \equiv \mathcal{H}^0_q \oplus \mathcal{H}^1_q \oplus \mathcal{H}_q^2 \oplus ...\oplus \mathcal{H}^n_q \oplus.....
\end{equation}
where $\mathcal{H}^0_q$ is the zero-particle space, the so called vacuum state and $\mathcal{H}^n_q$ is the $n$-particle space, with the super-script $n$ indicating the particle number. 
\subsection{Creation and Annihilation operators in Twisted basis}
We now introduce creation/annihilation operators for both twisted and quasi-commutative bases.

If we take the twisted symmetrized/anti-symmetrized momentum basis in this full quantum Hilbert space, we have identity operator in this basis as
\begin{equation}
\mathbf{I}_Q^\theta = \sum_{n=0}^\infty \frac{1}{n!}\int d^3 p_1 d^3 p_2....d^3p_n|\vec{p}_1,\vec{p}_2,....,\vec{p}_n)_\theta ~_\theta(\vec{p}_1,\vec{p}_2,....,\vec{p}_n|. 
\end{equation} The overlap of any two such basis states with different number of particles on $\mathcal{H}_Q$ vanishes:
\begin{equation}
_\theta(\vec{p}'_1,\vec{p}'_2,...\vec{p}'_N|\vec{p}_1, \vec{p}_2,...\vec{p}_M)_\theta = \delta_{NM}~ _\theta(\vec{p}'_1,\vec{p}'_2,...\vec{p}'_N|\vec{p}_1, \vec{p}_2,...\vec{p}_N)_\theta
\end{equation}
Then the creation and annihilation operators in this twisted momentum basis can be defined as
\begin{equation}
 \hat{a}^\ddag(\vec{p}) = \sum_{n=0}^\infty\frac{1}{n!} \int d^3p_1d^3p_2....d^3p_n~|\vec{p},\vec{p}_1, \vec{p}_2, ....,\vec{p}_n)_\theta~_\theta(\vec{p}_1,\vec{p}_2,....,\vec{p}_n| \label{top1}
\end{equation}
and
\begin{equation}
 \hat{a}(\vec{p}) = \sum_{n=0}^\infty\frac{1}{n!} \int d^3p_1d^3p_2....d^3p_n~|\vec{p}_1, \vec{p}_2, ....,\vec{p}_n)_\theta~_\theta(\vec{p}, \vec{p}_1,\vec{p}_2,....,\vec{p}_n|\label{top2}
\end{equation} 
Thus an arbitrary state on $\mathcal{H}_Q$ containing N-particles can be created by the N-fold action of the creation operators on the vacuum state as
\begin{equation}
 |\vec{p}_1,\vec{p}_2,...,\vec{p}_N)_\theta =\hat{a}^\ddag(\vec{p}_1)\hat{a}^\ddag(\vec{p}_2)....\hat{a}^\ddag(\vec{p}_N)|0)
\end{equation} and the further action of the creation operator on an arbitrary state can be defined as
\begin{equation}
 \hat{a}^\ddag(\vec{p})~|\vec{p}_1,\vec{p}_2,...\vec{p}_N)_\theta = |\vec{p},\vec{p}_1,\vec{p}_2,...,\vec{p}_N)_\theta~~~~~~~~~~~~~~~
\end{equation} Here, we take, by convention, that the creation operator creates the new particle at the first slot which is different from creating a particle at the last slot, unlike the commutative case - even for a twisted boson, as these field operators of twisted states obey the deformed commutation relations:
\begin{eqnarray}
 \hat{a}^\ddag(\vec{p})\hat{a}^\ddag(\vec{p}')&=& \eta~e^{ip'\wedge p}\hat{a}^\ddag(\vec{p}')\hat{a}^\ddag(\vec{p})\\
 \hat{a}(\vec{p})\hat{a}(\vec{p}')&=&\eta~e^{ip'\wedge p}\hat{a}(\vec{p}')\hat{a}(\vec{p})\\
 \text{and}~~~~~~~\hat{a}(\vec{p})\hat{a}^\ddag(\vec{p}')&=&\delta^3(\vec{p}-\vec{p}') +\eta~e^{ip\wedge p'}\hat{a}^\ddag(\vec{p}')\hat{a}(\vec{p}).
\end{eqnarray} On the other hand, the action of annihilation operator of the twisted state on an arbitrary state can be written as
\begin{eqnarray}
\nonumber \hat{a}(\vec{p})~|\vec{p}_1,\vec{p}_2,...,\vec{p}_N)_\theta &=& \sum_{a=0}^N \eta^{a-1}~e^{\frac{i}{2}(p\wedge p_1+...+p\wedge p_{a-1}+p\wedge p_{a+1}+...+p\wedge p_n)}\\
 & &  \delta^3(\vec{p}-\vec{p}_a)|\vec{p}_1,...\vec{p}_{a-1},\vec{p}_{a+1},...,\vec{p}_n)_\theta. 
\end{eqnarray}
\subsection{Creation and Annihilation operators in Quasi-commutative basis}
Now let us take the symmetrized/anti-symmetrized quasi-commutative orthonormal basis $|\vec{p}_1,\vec{p}_2,....,\vec{p}_n))$ for which we can define the identity operator on this full quantum Hilbert space $\mathcal{H}_Q$ as 
\begin{equation}
\mathbf{I}_Q = \sum_{n=0}^\infty \frac{1}{n!}\int d^3 p_1 d^3 p_2....d^3p_n|\vec{p}_1,\vec{p}_2,....,\vec{p}_n))((\vec{p}_1,\vec{p}_2,....,\vec{p}_n|. 
\end{equation}Again the overlap of any two states on $\mathcal{H}_Q$ with distinct number of particles vanishes:
\begin{equation}
_\pm((\vec{p}'_1,\vec{p}'_2,...\vec{p}'_N|\vec{p}_1, \vec{p}_2,...\vec{p}_M))_\pm = \delta_{NM}~ _\pm((\vec{p}'_1,\vec{p}'_2,...\vec{p}'_N|\vec{p}_1, \vec{p}_2,...\vec{p}_M))_\pm 
\end{equation}

We can now define analogously the creation and annihilation operators in the quasi-commutative basis as
\begin{equation}
 \hat{c}^\ddag(\vec{p}) = \sum_{n=0}^\infty\frac{1}{n!} \int d^3p_1d^3p_2....d^3p_n~|\vec{p},\vec{p}_1, \vec{p}_2, ....,\vec{p}_n))((\vec{p}_1,\vec{p}_2,....,\vec{p}_n| \label{qcop1}
\end{equation}
and
\begin{equation}
 \hat{c}(\vec{p}) = \sum_{n=0}^\infty\frac{1}{n!} \int d^3p_1d^3p_2....d^3p_n~|\vec{p}_1, \vec{p}_2, ....,\vec{p}_n))((\vec{p}, \vec{p}_1,\vec{p}_2,....,\vec{p}_n| \label{qcop2}
\end{equation}
where these new operators are related to the ones of twisted basis (\ref{top1}, \ref{top2} ) as
\begin{eqnarray}
 \hat{a}(\vec{p}) &=& \hat{c}(\vec{p}) e^{\frac{i}{2}p_i\theta^{ij}\hat{P}_j} \\
\text{and}~~~~~~~~ \hat{a}^\ddag(\vec{p}) &=&  e^{-\frac{i}{2}p_i\theta^{ij}\hat{P}_j} \hat{c}^\ddag(\vec{p})
\end{eqnarray} where $\hat{P}_j$ is the total momentum\footnote{Note that similar expression also occurs in \cite{b14}, but there $\hat{c}(\vec{p})$ and $\hat{c}^\ddag(\vec{p})$ stand for entirely commutative case ($\theta=0$), in contrast to ours, where $\theta$-dependence persists in their defining expressions (\ref{qcop1}, \ref{qcop2}) through the quasi-commutative basis. The fact that their (anti) commutation relations (\ref{qcom1},\ref{qcom3}) are just like their commutative ($\theta=0$) counterparts, which however develops $\theta$-deformation in the correlation function through deformed thermal wavelength in the more physical Voros basis, as will be shown subsequently, is the reason behind adopting the terminology \textquotedblleft quasi-commutative basis \textquotedblright.}. The similar actions of these new creation/annihilation operators are given by 

\begin{equation}
 |\vec{p}_1,\vec{p}_2,...,\vec{p}_N)) =\hat{c}^\ddag(\vec{p}_1)\hat{c}^\ddag(\vec{p}_2)....\hat{c}^\ddag(\vec{p}_N)|0)
\end{equation}

\begin{equation}
 \hat{c}^\ddag(\vec{p})~|\vec{p}_1,\vec{p}_2,...\vec{p}_N)) = |\vec{p},\vec{p}_1,\vec{p}_2,...,\vec{p}_N))~~~~~~~~~~~~~~~
\end{equation}
and
\begin{equation}
 \hat{c}(\vec{p})~|\vec{p}_1,\vec{p}_2,...,\vec{p}_N)) = \sum_{a=0}^N \eta^{a-1}\delta^3(\vec{p}-\vec{p}_a)|\vec{p}_1,...\vec{p}_{a-1},\vec{p}_{a+1},...,\vec{p}_n)).
\end{equation}
We can easily verify that these field operators obey the usual (i.e. like $\theta=0$) (anti) commutation relations:
\begin{eqnarray}
 \hat{c}^\ddag(\vec{p})\hat{c}^\ddag(\vec{p}')&=& \eta~\hat{c}^\ddag(\vec{p}')\hat{c}^\ddag(\vec{p})\label{qcom1}\\ 
 \hat{c}(\vec{p})\hat{c}(\vec{p}')&=&\eta~\hat{c}(\vec{p}')\hat{c}(\vec{p})\\
 \text{and}~~~~~~~\hat{c}(\vec{p})\hat{c}^\ddag(\vec{p}')&=&\delta^3(\vec{p}-\vec{p}') +\eta~\hat{c}^\ddag(\vec{p}')\hat{c}(\vec{p}). \label{qcom3}
\end{eqnarray}
\section{Field operators}
Now we can look for the abstract field operators without referring to any basis. Clearly, this is achieved by writing 
\begin{equation}
\hat{\Psi}\equiv\hat{\Psi}(\hat{\vec{x}})= \int d^3p ~ \hat{a}(\vec{p}) \otimes |\vec{p})
\end{equation} and 
\begin{equation}
\hat{\Psi}^{\ddag}\equiv\hat{\Psi}^{\ddag}(\hat{\vec{x}})= \int d^3p ~ \hat{a}^\ddag(\vec{p}) \otimes (\vec{p}|
\end{equation} 
Here, we should note that the first slot of tensor product is an operator acting on a particular quantum Hilbert space $\mathcal{H}^n_q$ to give an element of $\mathcal{H}^{n-1}_q$/$\mathcal{H}^{n+1}_q$ corresponding to one less (n-1)/one more (n+1) number of particles, while the second slot of the tensor product is the momentum eigenstate belonging to quantum Hilbert space $\mathcal{H}_q$, which incidentally is also an operator acting on classical Hilbert space (\ref{configuration}). Thus the field operators in position (Moyal/Voros basis) and momentum representations for the non-commutative case can then be understood as
\begin{equation}
\hat{\Psi}(\vec{x}_{M/V}) =  (1\otimes ~ _{M/V}(\vec{x}|)\hat{\Psi}(\hat{\vec{x}})= \int d^3p \hat{a}(\vec{p}) _{M/V}(\vec{x}|\vec{p}) 
\end{equation}
\begin{equation}
\hat{\Psi}^\ddag (\vec{x}_{M/V}) = \hat{\Psi}^{\ddag}(\hat{\vec{x}}) (1\otimes |\vec{x})_{M/V}) =\int d^3p \hat{a}^\ddag(\vec{p})(\vec{p}|\vec{x})_{M/V}
\end{equation}
and \begin{equation}
     \hat{\Psi}(\vec{p}) = (1\otimes(\vec{p}|)\hat{\Psi}(\hat{\vec{x}})=\int d^3p' \hat{a}(\vec{p}')\otimes (\vec{p}|\vec{p}')=\int d^3p' \hat{a}(\vec{p}')\delta^3(\vec{p}-\vec{p}')=\hat{a}(\vec{p})
    \end{equation}
Likewise, \begin{equation}
           \hat{\Psi}^\ddag(\vec{p}) = \hat{\Psi}^{\ddag}(\hat{\vec{x}}) (1\otimes |\vec{p})) = \hat{a}^\ddag(\vec{p})
          \end{equation}

And the action of the field operator on any arbitrary state of $\mathcal{H}_Q$ can be understood through the action of the first slot as   
\begin{equation}
\hat{\Psi} (~|p_1,... p_n ) \otimes 1~) = \int d^3p ~ \hat{a}(\vec{p}) |p_1,... p_n )\otimes |\vec{p} )
\end{equation}
\begin{equation}
\hat{\Psi}^\ddag (~|p_1,... p_n ) \otimes 1~) = \int d^3p ~ \hat{a}(\vec{p})^\ddag |p_1,... p_n )\otimes (\vec{p} |
\end{equation} so that we can understand the action of field operators on any state of $\mathcal{H}_Q$ in position and momentum representations as
\begin{eqnarray}
\hat{\Psi} (~|p_1,... p_n ) \otimes _{M/V}(\vec{x}|~) &=& \int d^3p ~ \hat{a}(\vec{p}) |p_1,... p_n )\otimes _{M/V}(\vec{x}|\vec{p} ) \nonumber \\
&=& \int d^3p _{M/V}(\vec{x}|\vec{p} ) \hat{a}(\vec{p}) |p_1,... p_n )
\end{eqnarray}
\begin{eqnarray}
\hat{\Psi}^\ddag (~|p_1,... p_n ) \otimes |\vec{x})_{M/V}~) &=& \int d^3p ~ \hat{a}^\ddag(\vec{p}) |p_1,... p_n )\otimes (\vec{p}|\vec{x} )_{M/V} \nonumber \\
&=& \int d^3p (\vec{p}|\vec{x} )_{M/V} \hat{a}^\ddag(\vec{p})|p_1,... p_n )
\end{eqnarray} and
\begin{eqnarray}
\nonumber \hat{\Psi} (~|p_1,... p_n ) \otimes (\vec{p}'|~) &=& \int d^3p ~ \hat{a}(\vec{p}) |p_1,... p_n )\otimes (\vec{p}'|\vec{p} )\\ \nonumber
 &=& \int d^3p ~ \hat{a}(\vec{p}) |p_1,... p_n ) \delta^3 (\vec{p}' - \vec{p})\\  
&=& \hat{a}(\vec{p}') |p_1,... p_n )
\end{eqnarray}
\begin{eqnarray}
\hat{\Psi}^\ddag (~|p_1,... p_n ) \otimes |p')~) &=& \int d^3p ~ \hat{a}(\vec{p})^\ddag |p_1,... p_n )\otimes (\vec{p}'|\vec{p} ) \nonumber \\
 &=& \hat{a}(\vec{p}') ^\ddag |p_1,... p_n )
\end{eqnarray}
Similarly for the new oscillators $\hat{c}(\vec{p})$ (\ref{qcop1}) and $\hat{c}^\ddag(\vec{p})$ (\ref{qcop2}), we have the field operators defined as 
\begin{equation}
 \hat{\Psi}_c\equiv \hat{\Psi}_c(\hat{\vec{x}}) = \int d^3p \hat{c}(\vec{p})\otimes |\vec{p}) 
\end{equation}
and \begin{equation}
     \hat{\Psi}_c^\ddag \equiv \hat{\Psi}_c^\ddag(\hat{\vec{x}}) = \int d^3p \hat{c}^\ddag(\vec{p})\otimes (\vec{p}|
    \end{equation}
with their actions defined in the analogous manner.
\section{Two-particle Correlation function}
Now we are interested to calculate the two-particle correlation function for a free gas in two and three dimensions using the canonical ensemble, i.e. to calculate the matrix elements $\frac{1}{Z}_{M/V}(\vec{r}_1,\vec{r}_2|e^{-\beta H}|\vec{r}_1,\vec{r}_2)_{M/V}$, where $Z$ is the canonical partition function and $H=\frac{1}{2m}(\vec{p}_1^2+\vec{p}_2^2)$ is the non-relativistic free particle Hamiltonian with $\beta=\frac{1}{k_BT}$. This has been already calculated in \cite{b11} in the twisted Moyal basis. However, since Moyal basis doesn't conform to the POVM unlike Voros \cite{b5}, our main interest will be to compute the resulting expression in Voros basis. Nevertheless we shall also present the corresponding computation in the Moyal basis to demonstrate their structural similarity, except for deformed thermal wavelength in the Voros case. Thus the violation of the Pauli principle seems to occur in either cases. However since the quasi-commutative basis, introduced in section VI.2 satisfy un-deformed 
exchange symmetry $\Sigma$ (\ref{exchangeformalism}) and orthonormality condition, unlike the twisted basis, we would like to re-analyze this in this quasi-commutative basis, to find out whether the Pauli principle, along with $SO(3)$ symmetry in 3D case, is restored or not.  
\subsection{Two-dimension}
\subsubsection{Twisted basis}
The twisted two-particle state in two-dimension is obtained (in the Voros/Moyal basis) by the 2-fold action of the field operator as,

\begin{equation}
 |\vec{r}_1,\vec{r}_2)_{\theta M/V} = \hat{\Psi}^\ddag(\vec{r}_1)_{M/V}\hat{\Psi}^\ddag(\vec{r}_2)_{M/V}|0)
 = \int d^2p_1d^2p_2(\vec{p}_1|\vec{r}_1)_{V/M}(\vec{p}_2|\vec{r}_2)_{V/M}\hat{a}^\ddag(\vec{p}_1)\hat{a}^\ddag(\vec{p}_2)|0) \label{TwistedMoyal}
\end{equation}
where in two dimension, it was shown in \cite{b5} 
\begin{equation}
_V(\vec{r}|\vec{p})= \sqrt{\frac{\theta}{2\pi}}e^{-\frac{\theta}{4}p^2}e^{i\vec{p}.\vec{r}}, ~~~~~_M(\vec{r}|\vec{p})=\frac{1}{2\pi}e^{i\vec{p}.\vec{r}} \label{overlaps}
\end{equation}
The two-particle correlation function in Moyal/Voros basis can therefore be written as 
\begin{equation}
 C_{\theta V/M}^{2D}(\vec{r}_1,\vec{r}_2) = \frac{1}{Z}~_{V/M\theta}(\vec{r}_1,\vec{r}_2|e^{-\beta H}|\vec{r}_1,\vec{r}_2)_{\theta V/M}
\end{equation} where Z stands for the partition function: \begin{equation}
                      Z = \int d^2r_1 d^2r_2~_{V/M\theta}(\vec{r}_1,\vec{r}_2|e^{-\beta H}|\vec{r}_1,\vec{r}_2)_{\theta V/M} 
                     \end{equation}
                     We can now write for the free particle Hamiltonian $H=\frac{1}{2m}\left(\vec{p}_1 ^2+ \vec{p}_2 ^2 \right)$, by inserting the 2-particle completeness relation satisfied by the twisted momentum basis $| \vec{k}_1,\vec{k}_2 )_\theta $ (\ref{twistedcommcompleteness}) 
                     \begin{equation}
                                   _{V/M\theta}(\vec{r}_1,\vec{r}_2|e^{-\beta H}|\vec{r}_1,\vec{r}_2)_{\theta V/M} = \int d^2k_1 d^2k_2 e^{-\frac{\beta}{2m}(k_1^2+k_2^2)}|_{V/M\theta}(\vec{r}_1,\vec{r}_2|\vec{k}_1,\vec{k}_2)_\theta |^2
                                  \end{equation}

Using (\ref{twistedorthonormality}) and (\ref{overlaps}), we get for the overlap of $| \vec{k}_1,\vec{k}_2 )$ with the Moyal basis (\ref{TwistedMoyal})
 \begin{equation}
 |_{M\theta}(\vec{r}_1,\vec{r}_2|\vec{k}_1,\vec{k}_2)_\theta |^2= \frac{1}{4(2\pi)^4}\left[2+\eta \big\{e^{i\theta_{ij} k_{2i} k_{1j}}e^{i(\vec{k}_1-\vec{k}_2).\vec{r}} + e^{-i\theta_{ij} k_{2i} k_{1j}}e^{-i(\vec{k}_1-\vec{k}_2).\vec{r}}\big\}\right]
 \end{equation}
  and like-wise for the Voros basis. Indeed, as it turns out, these two overlaps differs only by an exponential factors $e^{-\frac{\theta}{2}(k_1 ^2 + k_2 ^2)}$, apart from an unimportant factor of $(2\pi \theta)^2-$arising from the different normalization factors for Moyal and Voros basis (\ref{vorosbasis} , \ref{Moyal})
\begin{equation}
 |_{V\theta}(\vec{r}_1,\vec{r}_2|\vec{k}_1,\vec{k}_2)_\theta |^2=(2\pi\theta)^2e^{-\frac{\theta}{2}(k_1^2+k_2^2)} |_{M\theta}(\vec{r}_1,\vec{r}_2|\vec{k}_1,\vec{k}_2)_\theta |^2
\end{equation}

This indicates that the correlation function (in terms of the relative distance $\vec{r}=\vec{r}_1-\vec{r}_2$), computed in the Voros basis, will have the exactly same form as that of the Moyal basis, except that $\beta-$occurring in the correlation function in the Moyal basis. 
 \begin{equation}
      C_{\theta M}^{2D}(r) = \frac{1}{ A^2}\left[1+\eta\frac{1}{1+\frac{4\pi^2\theta^2}{\lambda^4}}e^{-\frac{2\pi}{\lambda^2(1+\frac{4\pi^2\theta^2}{\lambda^4})}r^2}\right],~~~~~\lambda=\sqrt{\frac{2\pi\beta}{m}}, \text{the mean thermal wavelength} \label{moyalian}
    \end{equation}
(which can be computed easily reproducing the expression derived already in \cite{b11}, (for a large area A)), has to be replaced by $\beta_{eff}=\beta+m\theta$. This finally yields for the correlation function in  Voros basis.

\begin{equation}
   C_{\theta V}^{2D}(r) = \frac{1}{A^2}\left[1+\eta \frac{1}{1+\frac{4\pi^2\theta^2}{\lambda_V^4}}e^{-\frac{2\pi}{\lambda_V^2\big(1+\frac{4\pi^2\theta^2}{\lambda_V^4}\big)}r^2}\right]\label{voros2}
 \end{equation}
where 
\begin{equation}
{\lambda^2_V} = \frac{2\pi\beta_\text{eff}}{m} = \lambda^2+2\pi\theta  \label{deformedlambda}
\end{equation}
 representing a non-commutative deformation of the mean thermal wavelength $\lambda = \sqrt{\frac{2\pi \beta}{m}}$ in the corresponding commutative ($\theta=0$) case \cite{b18}. Note that in contrast to Voros case, the \textquotedblleft Moyalian \textquotedblright~ expression (\ref{moyalian}) does not display any such deformation and therefore can be made arbitrarily small by taking arbitrarily large temperature $T=\frac{1}{\beta}$. On the other hand, the mean thermal wavelength in the Voros case can't be made smaller than $ \lesssim \sqrt{\theta}$ which is expected, as the Voros basis is supposed to capture the non-commutative feature correctly, by suppressing exponentially the oscillations with associated wavelength $ \lesssim \sqrt{\theta} $ in the corresponding wave function $\psi (z,\bar{z})\equiv  {_V(z|\psi)} $ \cite{b5}. Further, the relative distance $r=|\vec{r}_1-\vec{r}_2|$ occurring in the either expression (\ref{moyalian},\ref{voros2} ) can only be regarded as the true distance only in the Voros 
case, as the spectral distance \textit{a la} Connes could be computed in this case - unlike in the Moyal case \cite{b16}. And this difference stems from the fact that it is only the Voros basis that conforms to POVM-unlike the Moyal basis \cite{b5}.

Finally, we would like to say that this expression (\ref{voros2}) is different from the one obtained in \cite{b13} which was calculated with the aspects of braided twisted symmetry, see (\cite{b9}, \cite{b10}). As we had already mentioned that our approach is different from that of the braided twisted symmetry, it is quite obvious that we get different results.
\subsubsection{Quasi-commutative basis}
Now let us consider the usual symmetric/antisymmetric two-particle state in the quasi-commutative basis given by
\begin{equation}
 |\vec{r}_1,\vec{r}_2))_{V/M}= \hat{\Psi}_c^\ddag(\vec{r}_1)_{V/M} \hat{\Psi}_c^\ddag(\vec{r}_2)_{V/M} |0) = \int d^2p_1d^2p_2(\vec{p}_1|\vec{r}_1)_{V/M}(\vec{p}_2|\vec{r}_2)_{V/M}\hat{c}^\ddag(\vec{p}_1)\hat{c}^\ddag(\vec{p}_2)|0)
\end{equation}
Computing in the similar way as done in the twisted case, we finally obtain the two-particle correlation functions in the quasi-commutative bases as

\begin{equation}
    C_{cM}^{2D}(r)=\frac{1}{A^2}\left[1+ \eta e^{-\frac{2\pi}{\lambda^2}r^2}  \right] ~~~\text{and}~~~C_{cV}^{2D}(r)=\frac{1}{A^2}\left[1+\eta e^{-\frac{2\pi}{\lambda_V^2}r^2} \right]
\end{equation}
   The above expressions show that the two-particle correlation function in the quasi-commutative Moyal basis is just the same as the usual commutative result. However, in the quasi-commutative Voros basis, although it turns out to be structurally same with the commutative or quasi-commutative Moyal case, it picks up a non-commutative deformation through deformed mean thermal wavelength (\ref{deformedlambda}). Clearly, we recover the Pauli exclusion principle in both quasi-commutative Moyal and Voros cases.
\subsection{Three-dimension}
    In the same way, the two-particle correlation function in the twisted momentum basis in three-dimension can be carried out. The twisted two-particle Voros/Moyal basis is given by
    \begin{equation}
     |\vec{r}_1,\vec{r}_2)_{\theta V/M}= \int d^3p_1d^3p_2(\vec{p}_1|\vec{r}_1)_{V/M}(\vec{p}_2|\vec{r}_2)_{V/M} |\vec{p}_1,\vec{p}_2)_\theta
    \end{equation}
as in three-dimension, the overlap between the momentum basis and the Voros/Moyal basis are given by
\begin{equation}
  _V(\vec{r}|\vec{p})=\left(\frac{\theta}{2\pi}\right)^\frac{3}{4}e^{-\frac{\theta}{4}p^2}e^{i\vec{p}.\vec{r}}, ~~~~~ _M(\vec{r}|\vec{p})=\frac{1}{(2\pi)^\frac{3}{2}}e^{i\vec{p}.\vec{r}}
\end{equation}
We can calculate the correlation function in the same way as done in 2D case. But, as we had already mentioned that in three-dimension ~$\theta_{ij}$~is a 3x3 matrix antisymmetric matrix and ~$\vec{\theta}=\{\theta_k\}$~ is a vector dual to it. So it becomes more complicated to calculate the correlation function. However, we can simplify the calculation by taking the real vector ~$\vec{\theta}$~ along the third axis so that ~$\theta_1=\theta_2=0$~ and ~$\theta_3=\theta_{12}=\theta$.~With this we have, as in two-dimension,
\begin{equation}
  i\theta_{ij}k_{2i}k_{1j} =i\theta(k_{2x}k_{1y}-k_{2y}k_{1x})
\end{equation}
   
and, we can easily obtain the two-particle correlation function in three-dimension for the twisted Moyal and Voros bases, in the limit volume ~$V\rightarrow\infty$,~ as 
\begin{equation}
      C_{\theta M}^{3D}(r_\perp,r_\parallel)= \frac{1}{V^2}\left[1+\eta \frac{1}{1+\frac{4\pi^2\theta^2}{\lambda^4}}e^{-\Big\{\frac{2\pi}{\lambda^2\big(1+\frac{4\pi^2\theta^2}{\lambda^4}\big)}r_{\perp} ^2 +\frac{2\pi}{\lambda^2}r_{\parallel} ^2\Big\}}  \right] \label{corr3dm}
    \end{equation}and 
\begin{equation}
      C_{\theta V}^{3D}(r_\perp,r_\parallel)=\frac{1}{V^2}\left[1+\eta \frac{1}{1+\frac{4\pi^2\theta^2}{\lambda_V^4}}e^{-\frac{2\pi}{\lambda_V^2\big(1+\frac{4\pi^2\theta^2}{\lambda_V^4}\big)}r_{\perp} ^2-\frac{2\pi}{\lambda_V^2}r_{\parallel} ^2}\right],\label{corr3dv}
    \end{equation}
    where $r_{\perp} =\sqrt{r_x^2+r_y^2} $ and $r_{\parallel}  = r_z $ representing the relative separations along the transverse and longitudinal directions respectively, as determined by the $\vec{\theta}$ vector. Also, note that for a relative separation purely in the transverse direction i.e. $r_\perp\neq 0$ but $r_\parallel = 0$, both of these twisted expressions (\ref{corr3dm}-\ref{corr3dv}) go over to their appropriate 2D forms (\ref{moyalian}, \ref{voros2}).\\
    
The respective expressions in quasi-commutative basis take isotropic forms and are given by 
\begin{equation}
C_{cM}^{3D}(r)= \frac{1}{V^2}\left[1+\eta e^{-\frac{2\pi}{\lambda^2}r^2}  \right]~~~\text{and}~~~ C_{cV}^{3D}(r)= \frac{1}{V^2}\left[1+\eta e^{-\frac{2\pi}{\lambda^2_V}r^2} \right],
\end{equation} where $r=\sqrt{r_x^2+r_y^2+r_z^2}=\sqrt{r_{\perp}^2+r_{\parallel}^2}$.
Again note the structural similarity between Moyal and Voros cases, except that in Voros case, we will have to replace $\lambda\rightarrow\lambda_V$ (\ref{deformedlambda}) as before. Further, in the twisted basis one has both the symmetry breaking (from $SO(3)\rightarrow SO(2)$) as well as the violation of Pauli principle. In contrast, in the quasi-commutative basis we can preserve both $SO(3)$ symmetry and Pauli-principle. This breaking of $SO(3)$ symmetry in twisted case can therefore be attributed to the non-orthonormality condition of the twisted momentum basis which carries an extra phase factor depending on $\theta_{ij}$ in contrast to the quasi-commutative momentum basis which is orthonormal just like the commutative case.

Here in the above calculation we have chosen a specific form of ~$\theta_{ij}$~ which is indeed ~$\bar{\theta}_{ij}$~ where ~$\bar{\theta}_{ij}=(\bar{R}\theta\bar{R}^T)_{ij}$~ and ~$\bar{R}\in SO(3)$~ is the rotation in the configuration space such that ~$\hat{\bar{x}}_i = \bar{R}_{ij}\hat{x}_j$.~ This rotation in the configuration space will implement a unitary transformation on the quantum Hilbert space so that a one particle state will transform as 
\begin{equation}
 |\psi^{\bar{R}}) = U(\bar{R})|\psi)
\end{equation}
then the two-particle state will transform as (following (\ref{102}) in section \ref{sec:Two particle formalism}  )
\begin{equation}
|\psi_1)\otimes |\psi_2) \rightarrow |\psi_1^{\bar{R}})\otimes|\psi_2^{\bar{R}}) = \Delta_\theta(\bar{R})\big(|\psi_1)\otimes|\psi_2) \big)                                             
\end{equation}
where ~$\Delta_\theta(\bar{R})$~ is the deformed co-product (\ref{deformed co product}), given as ~$\Delta_\theta(\bar{R}) = F \Delta_0(\bar{R}) F^{-1}$ ~with ~$F = e^{\frac{i}{2}\theta_{ij}\hat{P}_i\otimes\hat{P}_j}$~ (\ref{twist}) being the abelian Drinfeld twist and ~$\Delta_0(\bar{R}) = U(\bar{R})\otimes  U(\bar{R})$ is the un-deformed co-product (\ref{undeformed product}).

With this, it is clear that we can write
\begin{equation}
 _{V/M\theta}(\vec{\bar{r}}_1,\vec{\bar{r}}_2|e^{-\beta H}|\vec{\bar{r}}_1,\vec{\bar{r}}_2)_{\theta V/M}=_{V/M\theta}(\vec{r}_1,\vec{r}_2|(F U^\ddag(\bar{R})\otimes U^\ddag(\bar{R})F^{-1})e^{-\beta H}(F U(\bar{R})\otimes U(\bar{R})F^{-1})|\vec{r}_1,\vec{r}_2)_{\theta V/M} \label{corr.}
\end{equation}
The Hamiltonian ~$H$~ in our case is the that of a pair of free particles, given by 
\begin{equation}
 H= \frac{1}{2m}(\vec{p}^2\otimes I +I\otimes\vec{p}^2)\Longrightarrow[H,F]=0. \label{commute}
\end{equation}
Thus, giving overall no effect, so the above equation (\ref{corr.}) will reduce to
\begin{equation}
  _{V/M\theta}(\vec{\bar{r}}_1,\vec{\bar{r}}_2|e^{-\beta H}|\vec{\bar{r}}_1,\vec{\bar{r}}_2)_{\theta V/M}=_{V/M\theta}(\vec{r}_1,\vec{r}_2|e^{-\beta H}|\vec{r}_1,\vec{r}_2)_{\theta V/M}
\end{equation}
This shows that although we had taken ~$\bar{\theta}_{ij}$~ to carry out the computation in the barred frame, the result will remain the same if we were to calculate in the fiducial frame taking ~$\theta_{ij}$.~ We made the choice of ~$\theta_{ij}$~ to make the calculation easier. However, it should be noted that this holds for the case of the free particle Hamiltonian only. This might not be true if the Hamiltonian have an interaction term, as (\ref{commute}) may not hold any more.

\subsection{Thermal Effective Potential}
  We can then compute the effective potential for each cases by putting the above expressions in the relation ~$ V(\vec{r})=-k_B T \ln C(\vec{r})$. For the convenience of comparison, we recast these expressions in terms of the dimensionless variables $(\frac{r_{\perp}}{\lambda})$, $(\frac{r_{\parallel}}{\lambda})$, $(\frac{r}{\lambda})$ and $(\frac{\theta}{\lambda^2})$, involving the un-deformed thermal wavelength $\lambda$ (\ref{moyalian}).

  For the twisted Moyal and Voros cases, we have the thermal effective potential as
\begin{equation}
     V^{2D}_{\theta M}(r) = -k_B T \ln C^{2D}_{\theta M} = -k_B T \ln\left[1+\eta \frac{1}{1+(2\pi\frac{\theta}{\lambda^2})^2}e^{-\frac{2\pi}{1+(2\pi\frac{\theta}{\lambda^2})^2}\frac{r^2}{\lambda^2}}\right]
    \end{equation}
and 
\begin{equation}
 V^{2D}_{\theta V}(r) = -k_B T \ln C^{2D}_{\theta V} = -k_B T \ln\left[1+\eta \frac{1}{\Big\{1+\frac{4\pi^2\frac{\theta^2}{\lambda^4}}{(1+2\pi\frac{\theta}{\lambda^2})^2}\Big\}} e^{-\frac{2\pi}{(1+2\pi\frac{\theta}{\lambda^2})\big\{1+\frac{4\pi^2\frac{\theta^2}{\lambda^4}}{(1+2\pi\frac{\theta}{\lambda^2})^2}\big\}}\frac{r^2}{\lambda^2}} \right]
\end{equation}
For the cases of quasi-commutative Moyal and Voros bases, in the same way, we have
\begin{equation}
V^{2D}_{c M}(r) = -k_B T \ln C^{2D}_{c M} = -k_B T \ln\left[1+ \eta e^{-2\pi\frac{r^2}{\lambda^2}}\right]
\end{equation}
and 
\begin{equation}
 V^{2D}_{c V}(r) = -k_B T \ln C^{2D}_{c V} = -k_B T \ln\left[1+\eta e^{-\frac{2\pi}{1+2\pi\frac{\theta}{\lambda^2}}\frac{r^2}{\lambda^2}} \right]
\end{equation}

We have the similar following expressions for the twisted Moyal and Voros cases in three-dimension, but now these depend both on $r_\perp$ and $r_\parallel$, as the $SO(3)$ symmetry is now broken to $SO(2)$:
 
\begin{equation}
V^{3D}_{\theta M}(r_\perp,r_\parallel) = -k_B T \ln C^{3D}_{\theta M} =-k_B T\ln\left[1+\eta \frac{1}{1+\frac{4\pi^2\theta^2}{\lambda^4}}e^{-\Big\{\frac{2\pi}{\big(1+\frac{4\pi^2\theta^2}{\lambda^4}\big)}\frac{r_{\perp}^2}{\lambda^2}  +2\pi\frac{r_{\parallel}^2}{\lambda^2} \Big\}}  \right]
\end{equation}
and
\begin{equation}
V^{3D}_{\theta V}(r_\perp,r_\parallel) = -k_B T \ln C^{3D}_{\theta V}=-k_B T \ln\left[1+\frac{1}{\Big\{1+\frac{4\pi^2\frac{\theta^2}{\lambda^4}}{(1+2\pi\frac{\theta}{\lambda^2})^2}\Big\}} e^{-\Big\{\frac{2\pi}{(1+2\pi\frac{\theta}{\lambda^2})\big\{1+\frac{4\pi^2\frac{\theta^2}{\lambda^4}}{(1+2\pi\frac{\theta}{\lambda^2})^2}\big\}}\frac{r_{\perp} ^2}{\lambda^2}+\frac{2\pi}{(1+2\pi\frac{\theta}{\lambda^2})}\frac{r_{\parallel} ^2}{\lambda^2}\Big\}}\right]
\end{equation}

and for the quasi-commutative Moyal and Voros bases, we have the same expression as in the two-dimension case, as $SO(3)$ symmetry is restored and the effective potential now depends only on $r$:  
\begin{equation}
V^{3D}_{c M}(r) = -k_B T \ln C^{3D}_{c M} = -k_B T \ln\left[1+ \eta e^{-2\pi\frac{r^2}{\lambda^2}}\right]
\end{equation}
and 
\begin{equation}
 V^{3D}_{c V}(r) = -k_B T \ln C^{3D}_{c V} = -k_B T \ln\left[1+\eta e^{-\frac{2\pi}{1+2\pi\frac{\theta}{\lambda^2}}\frac{r^2}{\lambda^2}} \right]
\end{equation}
These expressions are plotted in Fig. 1 (for 2D case) and in Fig. 2 (for 3D case).

\begin{figure}
\includegraphics[width=20cm,height=11cm]{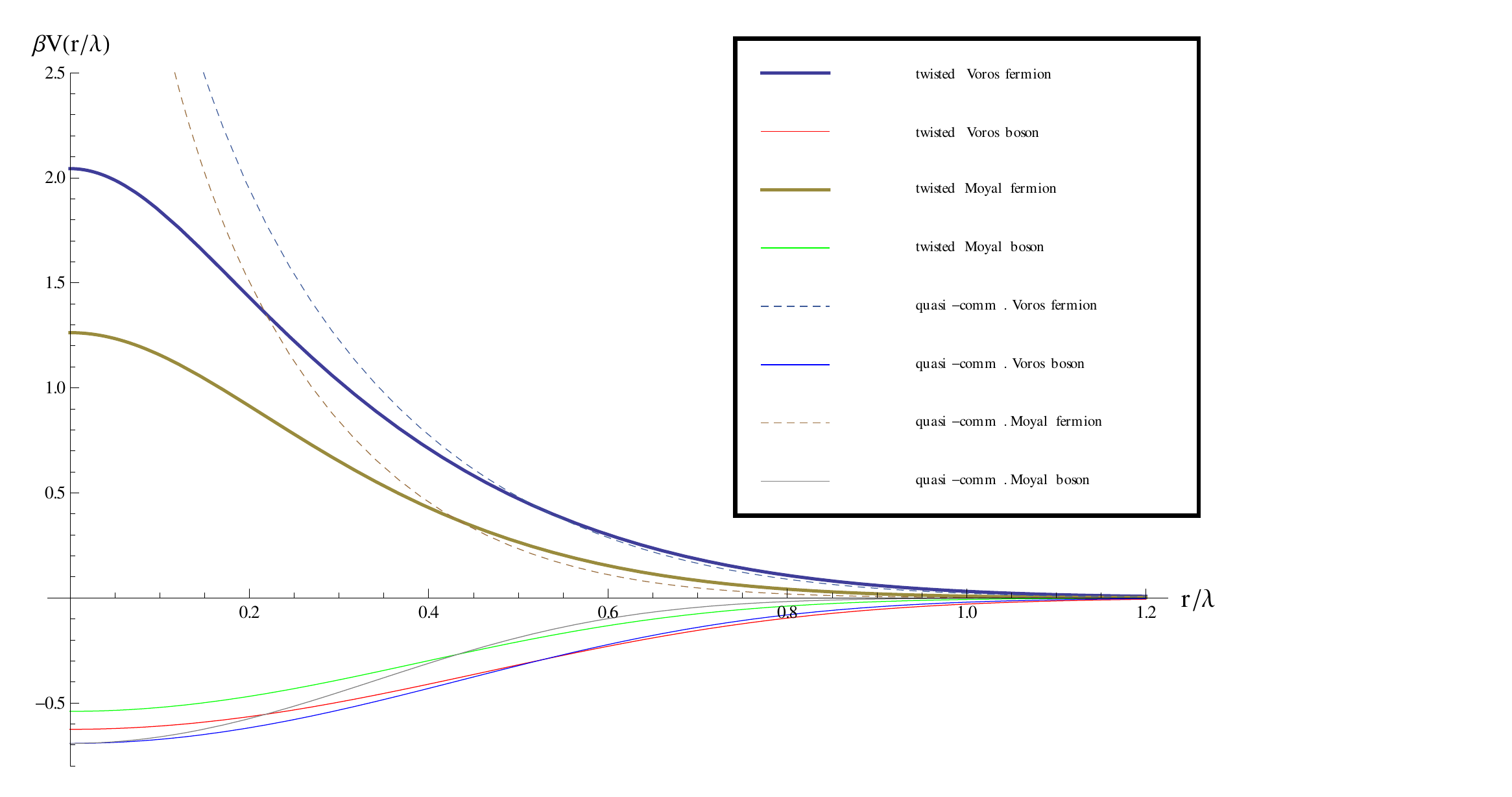}
\caption {\footnotesize \textbf{Thermal effective potential vs distance for different cases in two-dimension when $\frac{\theta}{\lambda^2}=0.1$}  }
\end{figure}

\begin{figure}
\includegraphics[width=8cm,height=5.5cm]{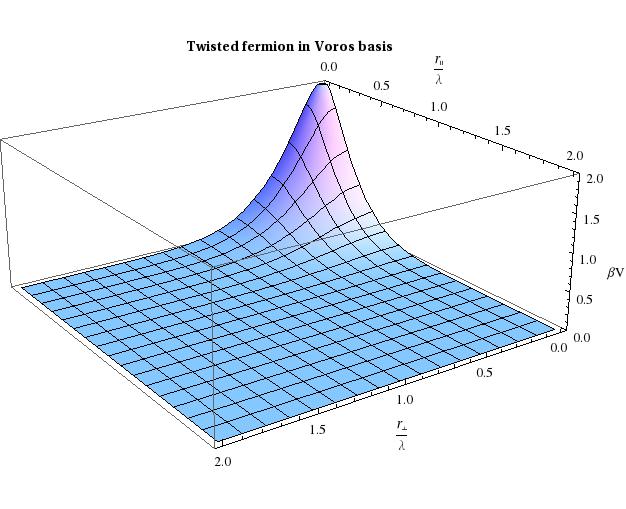}
\vspace{0.5cm}
\includegraphics[width=8cm,height=5.5cm]{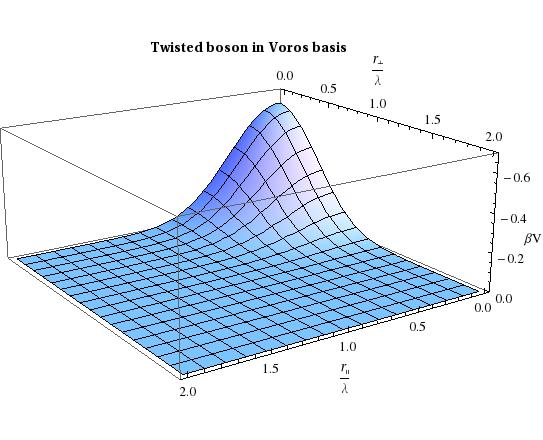}
\vspace{0.5cm}
\includegraphics[width=8cm,height=5.5cm]{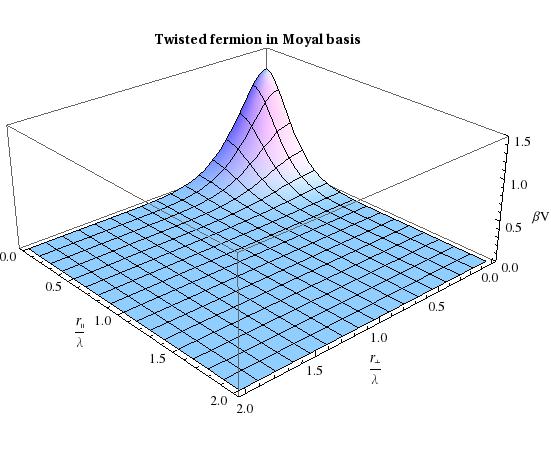}
\vspace{0.5cm}
\includegraphics[width=8cm,height=5.5cm]{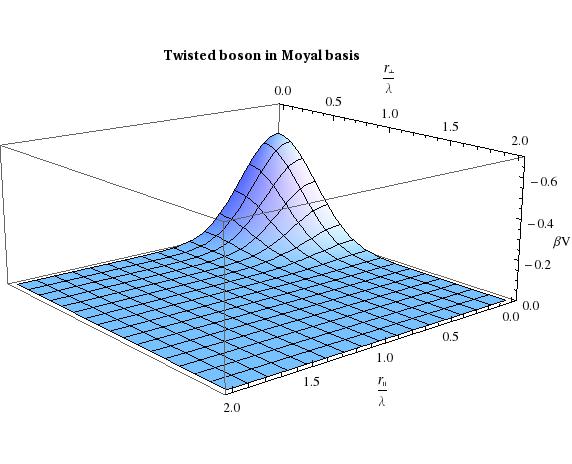}
\vspace{0.5cm}
\includegraphics[width=7cm,height=5cm]{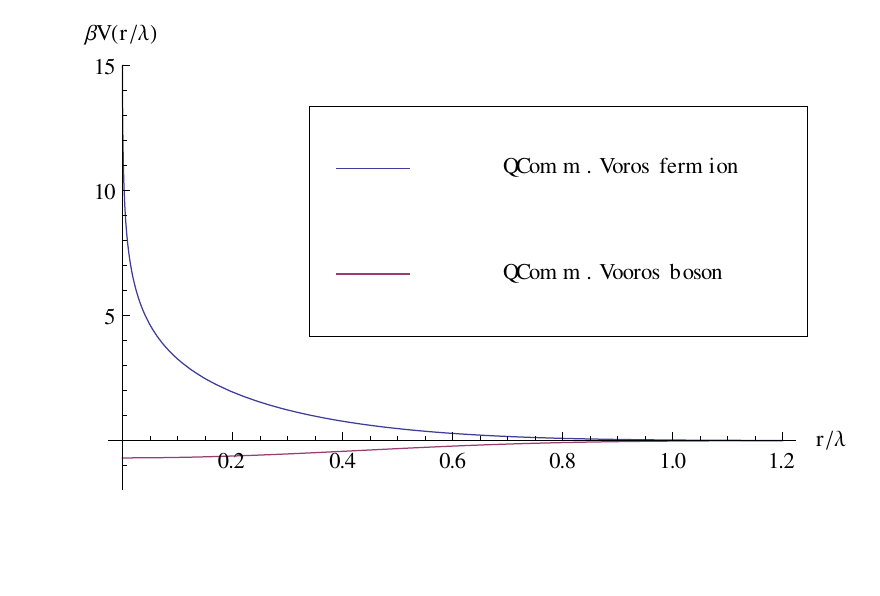}~~~~~~~~~~~~~~~~~~~~~~~~~~~~~~~~~~~~~
\vspace{0.5cm}
\includegraphics[width=7cm,height=5cm]{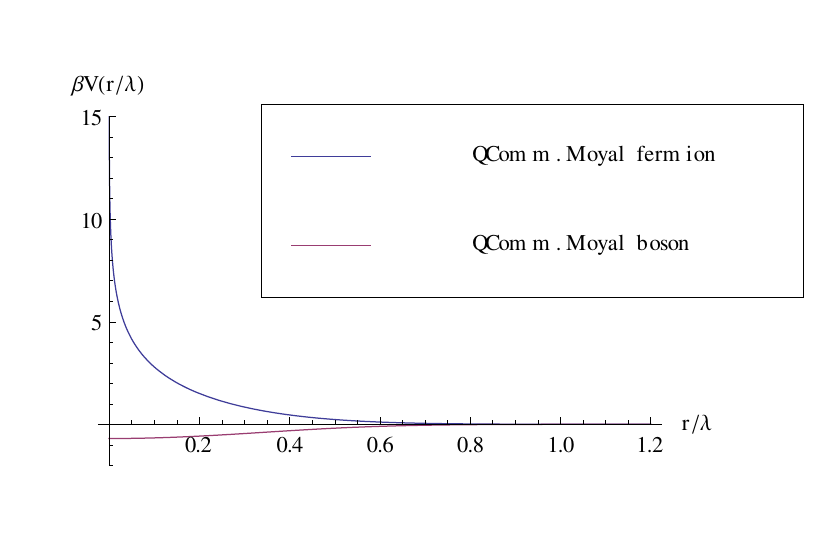}
\caption {\footnotesize \textbf{Thermal effective potential vs distance for $\frac{\theta}{\lambda^2}=0.1$ for each case in three-dimension. Note that, in the twisted case, this depends functionally on $r_{\perp}=\sqrt{r_x ^2 + r_y ^2}$ and $r_{\parallel}$, in contrast to quasi commutative case, where it depends only on $r$. }  }
\end{figure}

\newpage 
\section{Conclusion}
The issue of twisted symmetry in 2D/3D Noncommutative Moyal space has been re-visited in a completely operatorial framework using Hilbert-Schmidt operators to investigate whether the twisted bosons/fermions \cite{b14} necessarily occurs in conjunction with the twisted deformed coproduct in Moyal space \cite{b15}, where the twisted fermions were shown to violate Pauli principle \cite{b11}. Further, even within this scheme, it is shown that there exists a basis in the multi-particle sector called \textquotedblleft quasi-commutative basis\textquotedblright,~  which satisfies orthonormality and completeness relation  and is symmetric/antisymmetric under the usual i.e. un-deformed exchange operator, so that one has usual bosons/fermions and can avoid introducing twisted bosons/fermions. The correlation functions and the associated thermal effective potential is then shown to conform to Pauli principle, apart from preserving the $SO(3)$ symmetry in 3D case, both in Moyal and Voros basis, in contrast to the case of 
twisted bosons/fermions, where there is a $SO(3)\rightarrow SO(2)$ symmetry breaking. In all these cases, the resulting expressions in Moyal and Voros bases exhibit the same structural form, except that in the Voros case, one gets a $\theta$-deformed thermal wavelength ensuring that it has a non-vanishing lower bound, which is in conformity with the requirement that wavelengths $ \lesssim \sqrt \theta$ are suppressed exponentially. Thus, in Voros basis one gets a non-commutative deformation even in the quasi-commutative basis and this Voros basis should be regarded as physical as one can talk sensibly about the inter-particle separation, as one can introduce spectral distance a la Connes, unlike its \textquotedblleft Moyalian\textquotedblright~ counterpart \cite{b16}.

In this context, we would like to mention that the 3D Voros basis, which was introduced earlier in \cite{b6}, is shown here to saturate the 6D phase-space uncertainty by computing the variance matrix in a symplectic approach, although it does not correspond to maximally localised state in 3D space unlike its 2D counterpart. Besides, it has an isotropic structure, in the sense that the symplectic eigenvalues of the corresponding commutative variance matrix yield the same pair of eigenvalues for three independent \textquotedblleft modes\textquotedblright,~ which are now essentially decoupled from each other.

\pagebreak

\end{document}